\documentclass[superscriptaddress,amsmath,amssymb,aps,twocolumn,prl,longbibliography,nofootinbib
]{revtex4-2}
\usepackage{graphicx}
\usepackage{subfigure}
\usepackage{adjustbox}
\usepackage{bm}
\usepackage{color}
\usepackage{braket}
\usepackage{standalone}
\usepackage{multirow}
\usepackage{tikz}
\usepackage{mathrsfs}
\usepackage{dsfont}
\usepackage{comment}
\usepackage{amsmath}
\usepackage[colorlinks,bookmarks=true,citecolor=blue,linkcolor=red,urlcolor=blue]{hyperref}
\usepackage{cleveref}
\usepackage{orcidlink}
\renewcommand{\vec}[1]{\mbox{\boldmath$#1$}}

\newcommand{\RNum}[1]{\lowercase\expandafter{\romannumeral #1\relax}}

\graphicspath{{./figures/}}

\begin{document}

\title{Topologically protected emergent Fermi surface in an Abrikosov vortex lattice} 

\author{Songyang Pu}
\affiliation{Department of Physics and Astronomy, The University of Tennessee, Knoxville, TN 37996, USA}
\affiliation{Institute of Advanced Materials and Manufacturing, The University of Tennessee, Knoxville, TN 37920, USA}
\author{Jay D. Sau}
\affiliation{Condensed Matter Theory Center and Joint Quantum Institute, Department of Physics, University of Maryland,
College Park, Maryland 20742, USA}
\author{Rui-Xing Zhang}
\email{ruixing@utk.edu}
\affiliation{Department of Physics and Astronomy, The University of Tennessee, Knoxville, TN 37996, USA}
\affiliation{Institute of Advanced Materials and Manufacturing, The University of Tennessee, Knoxville, TN 37920, USA}
\affiliation{Department of Materials Science and Engineering, The University of Tennessee, Knoxville, TN 37996, USA}

\date{\today}

\begin{abstract}
We show that a three-dimensional (3D) fully gapped type-II superconductor can feature emergent in-gap Fermi surfaces of Caroli-de Gennes Matricon (CdGM) quasiparticles in the presence of an Abrikosov vortex lattice. In particular, these CdGM Fermi surfaces manifest in the emergent 3D band structure enabled by the intervortex tunneling physics, and their stability is guaranteed by a $\mathbb{Z}_2$ topological index. By developing an effective analytical theory, we find that each vortex line carrying a 1D nodal dispersion is a sufficient condition for the vortex lattice to form CdGM Fermi surfaces. Following this prediction, in-gap CdGM Fermi surfaces are numerically confirmed in a microscopic vortex-lattice simulation of a superconducting Dirac semimetal with an $s$-wave spin-singlet pairing, which is directly applicable to a large class of type-II superconductors such as LiFeAs. Remarkably, the CdGM Fermi surfaces persist even when the normal state is deformed to a doped insulator of trivial band topology. Our work establishes the vortex lattice as a new experimentally feasible control knob for emergent topological phenomena in conventional superconductors.         
\end{abstract}

\maketitle

{\it Introduction.---} Fermi-surface instability in metals accounts for many emergent quantum phenomena such as superconductivity~\cite{BCS1957} and charge density waves~\cite{gruner1988cdw}. For example, conventional superconductors (SCs) such as aluminum will experience a Cooper instability below the superconducting (SCing) critical temperature $T_c$, where electrons pair up to gap out the Fermi surface in an isotropic manner. Meanwhile, unconventional SCs such as the cuprates~\cite{bednorz1986possible} often feature an anisotropic pairing order that vanishes at points or lines in momentum space, with which the Fermi surface is partially gapped~\cite{sigrist1991rmp,tsuei2000cuprate,hirschfeld2011gap}.

While comprehending the fragility of various Fermi surfaces is crucial in solid-state systems, recent theoretical advances have suggested an intriguing possibility to topologically stabilize the Fermi surface in Bogoliubov-de Gennes (BdG) systems~\cite{zhao2013topo,kobayashi2014blount,zhao2016unified,Agterberg17,brydon2018BdG,santos2019pdw,link2020BdG,shaffer2020crystalline,setty2020BdG,dutta2021bdg,volkov2023twisted}. As noted in Ref.~\cite{Agterberg17}, a two-dimensional (2D) $\mathbb{Z}_2$ topological Fermi surface of BdG quasiparticles may exist in a centrosymmetric 3D SC if the pairing order spontaneously breaks the time-reversal symmetry (TRS). Despite the rapid theoretical developments, it was only until recently that evidence of BdG Fermi surface was reported in FeSe$_{1-x}$S$_x$~\cite{Nagashima22,Mizukami23,Matsuura23,Wu23}. In practice, the scarcity of known TRS-breaking unconventional SCs has significantly constrained the choice of candidate systems for BdG Fermi surfaces. 

In this work, we present a completely different mechanism for enabling $\mathbb{Z}_2$ topological Fermi surfaces in SCs, which only requires {\it s-wave spin-singlet} pairing and a field-induced Abrikosov vortex lattice~\cite{Abrikosov56,Franz00,Vafek01,chiu2015strongly,Liu15,chaudhary2020vortex,schirmer2022phase}. Distinct from the BdG Fermi surface, this novel type of topological Fermi surface describes an emergent metallicity of the Caroli-de Gennes Matricon (CdGM) vortex-bound states inside the bulk SCing gap~\cite{Caroli64}. In particular, we find that the CdGM Fermi surfaces necessarily show up in an ordered 2D lattice of 1D nodal vortices, which are known to exist in SCing Dirac semimetals (DSMs)~\cite{Qin19,Konig19,yan2020vortex,Hu22} and Luttinger semimetals~\cite{hu2023topological}. We have developed an effective theory for the proposed CdGM Fermi surface, which clarifies both its microscopic and topological origins. Numerically, we have explicitly confirmed the presence of CdGM Fermi surfaces in a SCing DSM system through a microscopic simulation of its vortex lattice. Remarkably, these zero-energy Fermi surfaces are found to persist even when the vortices are no longer nodal. Candidate materials and experimental consequences are also discussed.           

\begin{figure}
    \includegraphics[width=0.99\linewidth]{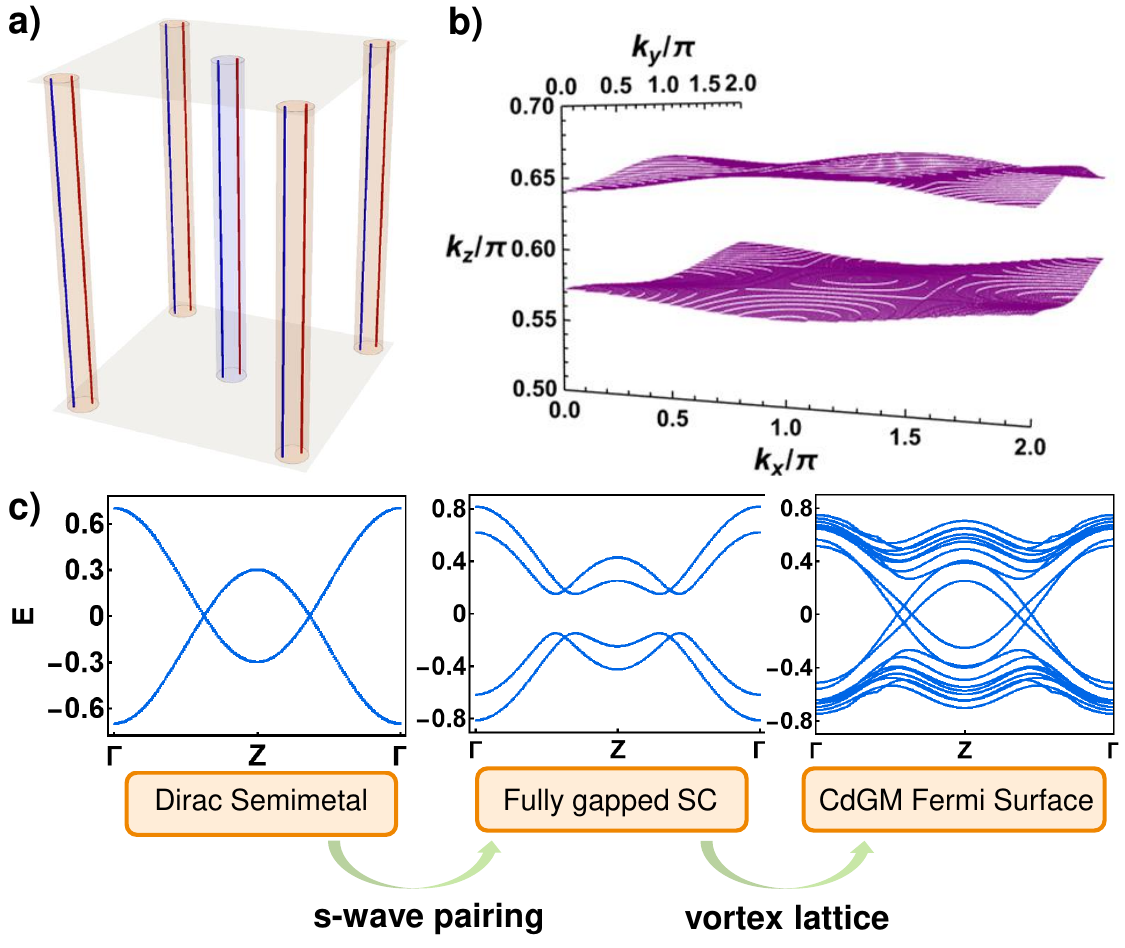}
    \caption{(a) A unit cell consists of two nodal vortices and each vortex has two gapless bound states with opposite angular momenta $l_z=\pm 1$. (b) The CdGM Fermi surfaces in the upper half Brillouin zone (BZ) $k_z>0$. Two other Fermi surfaces exist in the lower half BZ as a result of $z$-directional mirror symmetry. The Hamiltonian we used are the same as Fig.~\ref{fig3}(b-c) which are elucidated later in text. (c) A gapless DSM will develop a full SC gap under the $s$-wave pairing. In-gap CdGM Fermi surfaces will emerge when there is an Abrikosov vortex lattice.}
    \label{fig1}
\end{figure}

{\it A Lattice of Nodal Vortices.} --- We start with the Bogoliubov-de Gennes (BdG) theory of a 3D type-II superconductor with an s-wave spin-singlet pairing, where
\begin{equation}
    H({\bf k}) = \begin{pmatrix}
        h_0({\bf k}) - \mu & \Delta \\
        \Delta^* & \mu -h_0({\bf k}) \\
    \end{pmatrix}.
    \label{eq:BdG}
\end{equation}
Here $h_0({\bf k})$ denotes the normal-state Hamiltonian and $\mu$ is the chemical potential. The particle-hole symmetry (PHS) is $\Xi = \tau_y s_y {\cal K}$, where ${\cal K}$ is the complex conjugation operation and the Pauli matrices $\tau_{0,x,y,z}$ and $s_{0,x,y,z}$ are defined for the electron-hole and spin degrees of freedom, respectively. To describe a ${\bf B}$-field-induced vortex line, we consider a phase winding of the SC order parameter with $\Delta(r,\theta) = \Delta_0 \tanh(r/\xi_0) e^{i\theta}$, where $\xi_0$ is the SC coherence length and $(r,\theta)$ denote the in-plane polar coordinates. Hence, the vortex line manifests as an effective 1D class-D nanowire and it can further trap low-energy CdGM bound states that disperse along $k_z$. 

Should the vortex line respect an $n$-fold rotation symmetry around $z$, the CdGM modes are further labeled by a $z$-directional angular momentum $l_z\in\mathbb{Z}$ (mod $n$), whose value will be flipped under the PHS operation. As a consequence, whenever two PHS-related CdGM bands cross at zero energy, the band crossing is protected by $C_n$ if $l_z\notin\{0,\frac{n}{2}\}$~\cite{hu2023topological}. This ``worm-hole"-like mechanism allows for a 1D gapless {\it nodal vortex} state to emerge in a 3D fully gapped superconductor. Notably, the above definition of a nodal vortex has assumed a single-vortex limit with the vortex lattice constant $L\gg \xi_0$. As required by the flux quantization condition, an increase ${\bf B}$ necessarily reduces $L$, further promoting the inter-vortex physics that was ignored previously. We thus expect a nodal vortex lattice to display an emergent 3D band structure, the nature of which will be the focus of this work.     

\begin{figure}[t]
    \centering    
    \includegraphics[width=\linewidth]{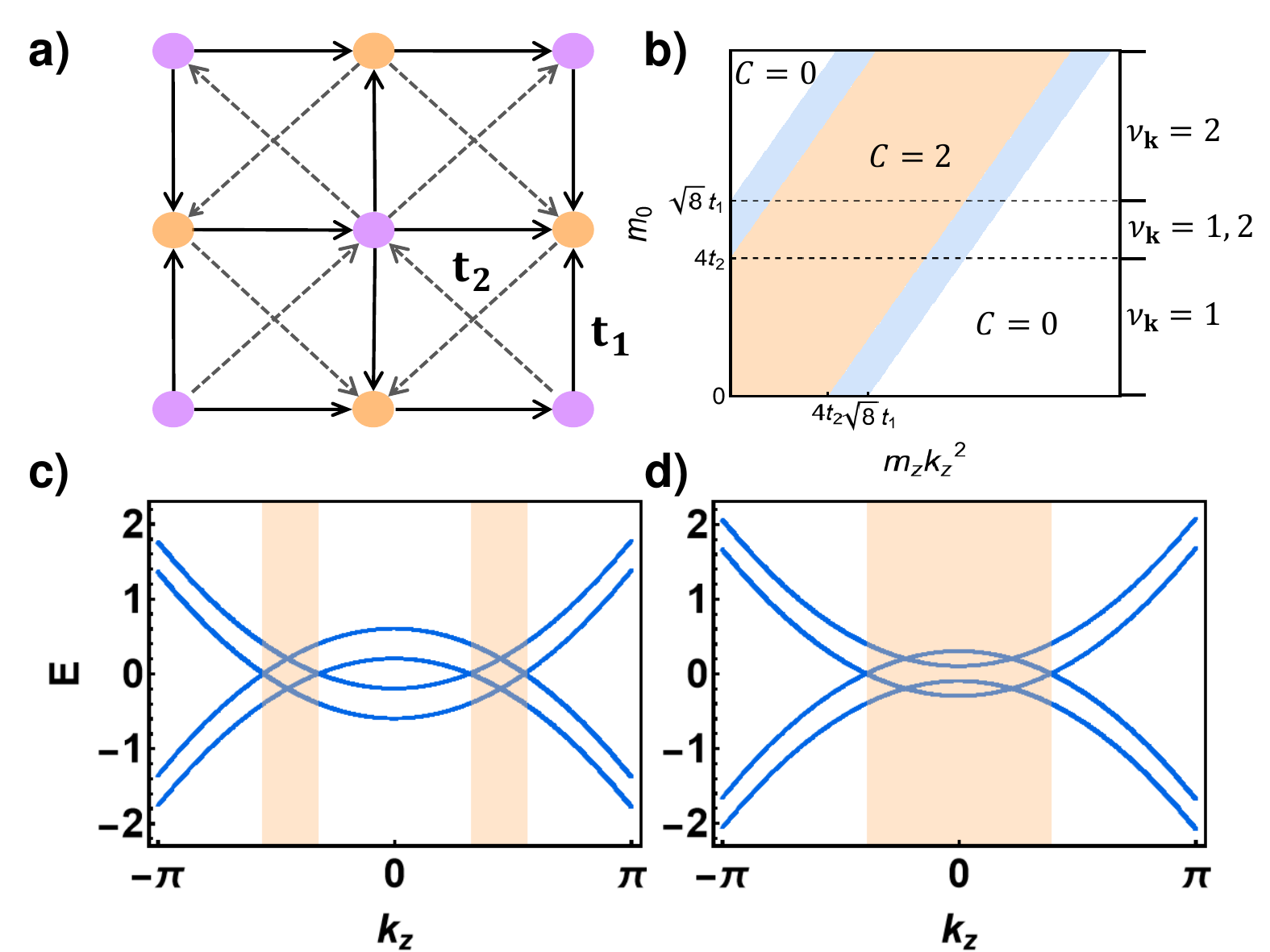}
    \caption{
    (a) Schematic of in-plane hopping events between neighboring vortices in the effective model ${\cal H}({\bf k)}$. The phase $\zeta_{\mathbf{R}_1,\mathbf{R}_2}$ is $\pi/2$ along the direction of arrows and $-\pi/2$ if otherwise. (b) Phase diagram of ${\cal H}({\bf k})$. The blue regions mark the range of $k_z$ where CdGM Fermi surfaces emerge. For $t_2=t_1 /2$ and $m_z=2t_1$, band dispersions along $k_z$ for $m_1=4t_1$ (with $\nu_{\bar{\Gamma}}=2$) and $m_0=t_1$ (with $\nu_{\bar{\Gamma}}=1$) are shown in (c) and (d), respectively. }
    \label{fig2}
\end{figure}

Without loss of generality, we focus on a square lattice of 1D nodal vortices that is schematically shown in Fig.~\ref{fig2} (a). Since each vortex line encloses a $\pi$ flux quantum, a primitive cell of the vortex lattice consists of two inequivalent vortices, dubbed $A$ and $B$. In our setup, we place the two inequivalent vortex cores at ${\bf R}_{A}=\frac{L}{4}(1,1)$ and ${\bf R}_{B}=\frac{3L}{4}(1,1)$, respectively. Therefore, the net phase pattern of $\Delta({\bf r})$ is given by $\theta({\bf r})=\theta_A({\bf r})+\theta_B({\bf r})$, where $\theta_\alpha =\sum_i {\rm arg}({\bf r}-{\bf R}_{i\alpha})$ arises from the vortex $\alpha$ at ${\bf R}_i$.

Further considering the magnetic gauge field ${\bf A(r)}$, a CdGM quasiparticle generally experiences an effective vector potential ${\bf \Omega(r)}$, with ${\bf \Omega}=\left(\vec{\Omega}_A + \vec{\Omega}_B\right)/2$ and $ \vec{\Omega}_\alpha=\nabla\theta_\alpha-{e\over \hbar c}{\bf A}$~\cite{Franz00}. In particular, ${\bf \Omega}$ can be solved together with the London equations~\cite{Vafek01,chiu2015strongly}, where
\begin{equation}
    \vec{\Omega}_\alpha({\bf r})=2\pi\lambda^2\int{d^2k\over (2\pi)^2}{i {\bf k}\times \hat{z}\over 1+\lambda^2|{\bf k}|^2}\sum_ie^{i{\bf k}\cdot ({\bf r}-{\bf R}_{i\alpha})}
    \label{eq:Omega_a}
\end{equation}
Here $\lambda$ is the London penetration length and we choose $\lambda=10L$.
When a quasiparticle hops from ${\bf r}_1$ to ${\bf r}_2$, the hopping coefficient $t_{\bf r_1,r_2}$ is proportional to $e^{i\zeta_{\mathbf{r}_1,\mathbf{r}_2}}$ following the Pierels substitution, where the displacement-dependent phase factor $\zeta_{\mathbf{r}_1,\mathbf{r}_2}=\int_{\mathbf{r}_1}^{\mathbf{r}_2}\vec{\Omega}\cdot d\vec{l}$. Therefore, the vortex lattice system can be viewed as a Hofstadter problem~\cite{hofstadter1976} with an effective magnetic field ${\cal B}=\nabla \times {\bf \Omega}$, which we will solve both analytically and numerically.    

{\it $\mathbb{Z}_2$ Topological CdGM Fermi Surface} --- Based on the gauge structure in Eq.~\ref{eq:Omega_a}, in the Supplemental Material (SM)~\cite{SM}, we analytically find that the hopping phase $\zeta_{\mathbf{R}_1,\mathbf{R}_2}$ between any two vortices at ${{\bf R}_{1,2}}$ is $\pm \frac{\pi}{2}$, implying that $t_{{\bf R}_1,{\bf R}_2}$ should be purely imaginary (i.e., $\propto \pm i$). This motivates us to construct an effective minimal tight-binding model for the array of 1D nodal vortices. Our choice of basis is $\Psi=(|A,+\rangle, |B,+\rangle, |A,-\rangle, |B,-\rangle)$, where $\pm$ is short for $l_z=\pm 1$. This leads to a $k$-space Hamiltonian,
\begin{equation}
    {\cal H}(\vec{k})= g({\bf k}_\parallel)\tilde{\tau}_0\otimes \sigma_0 +m(k_z)\tilde{\tau}_z \otimes \sigma_0,
    \label{H-TB}
\end{equation}
where $\tilde{\tau}_i$ and $\sigma_i$ are Pauli matrices for $l_z$ and sub-lattice degrees of freedom respectively. Here, $m(k_z)=(m_0-m_zk_z^2)$ with $m_0m_z>0$ describes the gapless $k_z$ dispersion of each nodal vortex. ${\cal H}(\vec{k})$ respects the four-fold rotation symmetry $C_4$, the inversion symmetry ${\cal P}$ and the PHS $\Xi$, whose explicit forms are discussed in the SM~\cite{SM}. As for the in-plane part $g({\bf k}_\parallel)$, we consider hopping events within the same $l_z$ sector among the nearest-neighboring ($t_1$) and next-nearest-neighboring ($t_2$) vortices. For simplicity, we have ignored the $l_z$-flipping terms, which will only quantitatively, but not qualitatively, change our target physics. In particular, we follow the gauge convention in Fig.~\ref{fig2} (a) and define $\zeta_{\mathbf{R}_1,\mathbf{R}_2}$ to be $\pi/2$ if the particle hops along the arrow, and $-\pi/2$ if otherwise. This immediately leads to $g({\bf k}_\parallel)=2t_1\sin {k_x+k_y\over 2}\sigma_x-2t_1\cos {k_x-k_y\over 2}\sigma_y+2t_2(\sin k_y-\sin k_x)\sigma_z$. Notably, $g({\bf k}_\parallel)$ itself reproduces the Hamiltonian of a 2D vortex Majorana lattice in Ref.~\cite{Liu15} which shares the same $\mathbb{Z}_2$ gauge structure, even though the CdGM states in our 3D vortex-lattice system have no Majorana nature. 

When viewing $k_z$ as a tunable parameter, we obtain a rather intuitive interpretation of ${\cal H}({\bf k})$: 
\begin{itemize}
    \item ${\cal H}_0({\bf k}_\parallel)=g({\bf k}_\parallel) \tilde{\tau}_0$ is a 2D class-D topologically gapped phase with a BdG Chern number $|C|=2$.
    \item $m(k_z)$ acts as an effective chemical potential for ${\cal H}_0$.  
\end{itemize}
The second point immediately follows the matrix structure of ${\cal H}$. As for ${\cal H}_0({\bf k}_\parallel)$, we note that each $g({\bf k}_\parallel)$ is essentially a square-lattice analog of the Haldane model~\cite{Haldane88}, which thus explains the origin of a non-zero $C$. Starting with $m(k_z)=0$, the effective Fermi level initially lies inside the topological gap of ${\cal H}_0$. As we gradually tune $k_z$ or $m(k_z)$, the Fermi level starts to cross either the conduction or valence Chern bands, leading to a metallic state with a 1D Fermi loop at $E=0$. After completely passing through the Chern bands, the 2D system recovers an energy gap. If we switch back to the 3D perspective, the 1D Fermi loops in the $k_x-k_y$ plane are smoothly stacked into a 2D Fermi surface in the 3D BZ while tuning $k_z$. We thus arrive at a striking conclusion that {\it an ordered lattice of nodal vortices will form a Fermi surface of zero-energy CdGM quasiparticles}.          

We now provide a quantitative understanding of the CdGM Fermi surface defined above. For the conduction Chern band, it is easy to show that the band top occurs at  $(k_x,k_y)=(\pm \pi/2,\pm \pi/2)$ with $E_{\rm max}=2\sqrt{2}t_1$, while the band bottom happens at $(k_x,k_y)=(\pm \pi/2,\mp \pi/2)$ with $E_{\rm min}=4t_2$ when assuming $t_1\geq 2t_2$. The competition between the effective Fermi level $m(k_z)$ and the bandwidth of the Chern band is well-captured by the phase diagram in Fig.~\ref{fig2}(b). Specifically, the orange region depicts the choices of $(m_0, m_z, k_z)$ with which the Fermi level lives inside the Chern gap. The blue and white regions focus on the scenarios when the Chern band is partially and fully filled, respectively. The number of CdGM Fermi surfaces for $k_z>0$ can be read out by drawing a horizontal reference line ${\cal L}$ in the phase diagram and counting the number of crossings between ${\cal L}$ and the blue regions. Note that there will be an equal number of CdGM Fermi surfaces for $k_z<0$ thanks to the mirror symmetry $M_z$. 

However, this counting strategy can be tricky for $4t_2 < m_0<2\sqrt{2} t_1$, when the two CdGM Fermi surfaces near $k_z=0$ start to merge and get reconstructed. As an alternative approach, we define $\nu_{{\bf k}_\parallel}\in \mathbb{Z}_{\geq 0}$ as the number of zero-energy band crossings along $k_z>0$ for any fixed in-plane ${\bf k}_\parallel$. As a concrete example, we plot the $k_z$-dispersions at $\bar{\Gamma}=(0,0)$ for $\nu_{\bar{\Gamma}}=2$ and $\nu_{\bar{\Gamma}}=1$ in Figs.~\ref{fig2} (c) and (d), respectively. For any ${\bf k}_\parallel$, we find that (i) $\nu_{{\bf k}_\parallel}=1$ when $m_0<4 t_2$ and (ii) $\nu_{{\bf k}_\parallel}=2$ when $m_0>2\sqrt{2} t_1$, just as we have expected. As for $4t_2 < m_0<2\sqrt{2} t_1$, $\nu_{{\bf k}_\parallel}\in \{1,2\}$ is found to be ${\bf k}_\parallel$ dependent due to the complex Fermi-surface topology. 

\begin{figure*}[t]
    \includegraphics[width=0.95\textwidth]{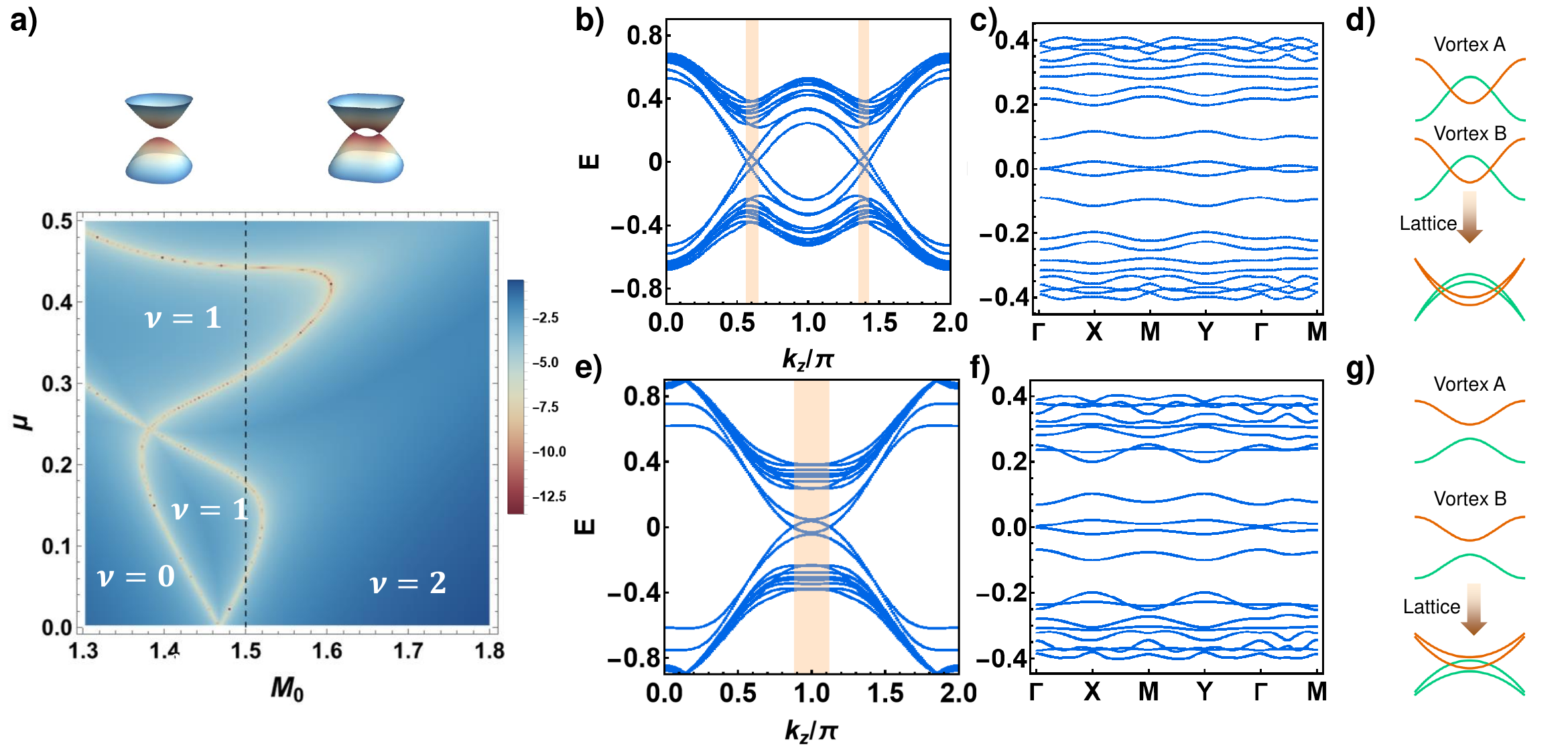}
    \caption{(a) Vortex-lattice topological phase diagram for a superconducting Dirac semimetal as a function of $\mu$ and $M_0$. The color shows the logarithm of the energy gap at $\vec{k}=(0,0,\pi)$ and $\nu$ is the number of CdGM Ferm surfaces for the half BZ with $k_z>0$. The black dashed reference line marks the normal-state phase boundary between a DSM and a topologically trivial insulator or metal. For $\nu=2$ phase, the out-of-plane dispersion along $k_z$ and the in-plane energy dispersion at $k_z=1.36\pi$ are shown in (b) and (c), with $\mu=0.1$, $M_0=1.8$ (i.e., the normal state is a doped DSM). The orange highlights the range of $k_z$ where the in-plane bands carry a $|C|=2$. For $\nu=1$ phase, the out-of-plane dispersion along $k_z$ and the in-plane energy dispersion at $k_z=0.88\pi$ are shown in (e) and (f), with $\mu=0.1$, $M_0=1.45$ (i.e., the normal state is a trivial metal). The orange highlights the range of $k_z$ where the in-plane bands carry a $|C|=1$. We choose $M_1=v=2M_2=1$ and $\Delta_0=0.15$ for all numerical simulations. We schematically illustrate how the hybridization between vortices A and B within the same unit cell generates the CdGM Fermi surfaces for $\nu=2$ in (d) and $\nu=1$ in (g).
    }
    \label{fig3}
\end{figure*}

Are the CdGM Fermi surfaces stable? Assuming our target system is centrosymmetric, the product of inversion symmetry ${\cal P}$ and the PHS $\Xi$ defines an effective 0D PHS ${\cal W}$, which transforms the Hamiltonian as ${\cal W}{\cal H}({\bf k}){\cal W}^{-1} = -{\cal H}^T({\bf k})$. This symmetry enables an antisymmetrization of the original Hamiltonian as $\tilde{\cal H}({\bf k}) = \Omega {\cal H}({\bf k}) \Omega^\dagger$, with $\tilde{\cal H}^T({\bf k})=-\tilde{\cal H}({\bf k})$ and $\Omega$ an unitary matrix~\cite{Agterberg17}. A Pfaffian ${\rm Pf}[\tilde{\cal H}(\vec{k})]$ can thus be defined for every ${\bf k}$. As a result, one can define a $\mathbb{Z}_2$ topological charge $\eta$ for a 3D Fermi surface $\Sigma$ as $(-1)^\eta={\rm sgn}\left[{\rm Pf}[\tilde{H}({\bf k}_1)]{\rm Pf}[\tilde{H}({\bf k}_2)]\right]$, where ${\bf k}_{1,2}$ are on different sides of $\Sigma$. When $\eta=1$, a sign flip of the Pfaffian occurs across $\Sigma$, which necessarily forces the energy gap to close at zero energy, leading to a topologically protected Fermi surface at $E=0$~\cite{kobayashi2014blount,zhao2016unified}. As we have confirmed numerically, all our CdGM Fermi surfaces have $\eta=1$ and their stability is hence guaranteed by the ${\cal P}\Xi$-protected $\mathbb{Z}_2$ topology.

We hope to highlight that our theory of CdGM Fermi surface only requires a 3D inversion symmetry and the built-in PHS, which can thus be applied to $C_2$, $C_4$, and $C_6$-invariant vortex geometry. For example, to describe a real-world vortex lattice that is often triangular, an effective model similar to Eq.~\ref{H-TB} can be constructed analogously by replacing $g({\bf k}_\parallel)$ with a triangular-lattice analog of the Haldane model~\cite{Liu15}. Nonetheless, the relation between the nodal vortices and the CdGM Fermi surface is general and independent of the explicit vortex geometry.  

{\it Numerical Proof of CdGM Fermi Surfaces.---} To further justify our theory of CdGM Fermi surfaces, we now perform a microscopic vortex-lattice simulation of an s-wave spin-singlet SC, whose normal state realizes a 3D Dirac semimetal (DSM)~\cite{wang2012dirac}. As discussed in Ref.~\cite{Qin19,Konig19,yan2020vortex,Hu22}, the vortex line of a SCing DSM features a pair of CdGM band crossings between the $l_z=\pm 1$ branches. Therefore, CdGM Fermi surfaces should emerge in a DSM-based SC when a vortex lattice is turned on.  

Our choice of normal-state Hamiltonian is $h_0({\bf k})=M({\bf k})s_0\otimes \kappa_z + v (\sin k_x s_z\otimes \kappa_x - \sin k_ys_0\otimes \kappa_y)$, where $M({\bf k})=M_0 - M_1(\cos k_x + \cos k_y) - M_2\cos k_z$ and $\kappa_i$ are the Pauli matrices for orbital degrees of freedom. When $M_0>2M_1-M_2$, $h_0({\bf k})$ achieves a 3D DSM phase with a pair of $C_4$-protected Dirac nodes around $Z=(0,0,\pi)$, as shown in Fig.~\ref{fig1} (c). Both the SC and the vortex can be implemented following the BdG formalism in Eq.~\ref{eq:BdG}. In the SM~\cite{SM}, we analytically solve for its vortex-bound state in the continuum limit, confirming the nodal vortex phase in the single-vortex limit. The 3D band structure of a square lattice of vortices can be numerically simulated following the methodology developed in Ref.~\cite{Franz00}. We have chosen the vortex lattice constant $L=14 a_0$, where $a_0=1$ denotes the atomic lattice constant. More simulation details can be found in the SM~\cite{SM}. 

The topological feature of the vortex-lattice bands can be studied by tracing the gap closing conditions at the high-symmetry momenta, which sufficiently track both the creation and annihilation of a CdGM Fermi surface. In Fig.~\ref{fig3} (a), we numerically map out the energy gap at $Z$ as a function of $M_0$ and the chemical potential $\mu$, where the gap-closing lines separate the phase diagram into multiple patches. We further exploit $\nu=\nu_{\bar{\Gamma}}$ defined earlier as an effective counting of the CdGM Fermi surfaces in the half BZ. As an example, Fig.~\ref{fig3} (b) shows the $k_z$ dispersion of a $\nu=2$ phase at $k_x=k_y=0$, where $M_0=1.8$ and $\mu=0.1$. The in-plane dispersion at a fixed $k_z=1.36\pi$ is further shown in Fig.~\ref{fig3} (c). Remarkably, when the normal state is a DSM (i.e., when $M_0>1.5$), we always find $\nu=2$, implying the existence of two CdGM Fermi surfaces in the half BZ. These CdGM Fermi surfaces are explicitly visualized in Fig.~\ref{fig1} (b) by systematically searching for zero-energy band crossings in the entire 3D BZ. Furthermore, their topological nature has also been numerically confirmed by calculating the $\mathbb{Z}_2$ topological charge $\eta$.          

Surprisingly, Fig.~\ref{fig3} (a) also reveals a $\nu=1$ phase for $M_0<2M_1-M_2$, when the normal state is a doped {\it topological trivial} insulator and every vortex is gapped by itself. In this case, we numerically confirm the existence of one CdGM Fermi surface in the half BZ by plotting both the out-of-plane and in-plane dispersions in Figs.~\ref{fig3} (e) and (f), respectively. This exotic phenomenon can be understood via a simple vortex hybridization picture between the two vortices A and B within the same unit cell. To facilitate the discussion, let us focus on the $k_z$ dispersions along the rotation axis $\bar{\Gamma}$, where each CdGM band is labeled by $l_z=1$ (in orange) and $l_z=-1$ (in green). Crucially, hybridization-induced level repulsion will only occur between CdGM bands with the same $l_z$. Denoting this effect of level repulsion as $\delta E$, zero-energy band crossings will occur when its strength exceeds the original gap of each vortex, as schematically shown in Fig.~\ref{fig3} (g). Away from $\bar{\Gamma}$, the zero-energy crossings will be connected to form a closed Fermi surface, leading to a $\nu=1$ phase. As shown in Fig.~\ref{fig3} (d), one can carry out a similar analysis to interpret the $\nu=2$ phase in Fig.~\ref{fig3} (b). In the SM~\cite{SM}, we numerically evaluate $\delta E$ as a function of the lattice constant $L$, which fits well to a simple exponential function $\delta E=C\Delta_0\exp[-L/L_0]$ with $C=2.28$ and $L_0 = 10.20$. Since the value of $L$ is controlled by ${\bf B}$ in experiments, one can always tune ${\bf B}$ to make $\delta E$ large enough to overcome the single-vortex gap. Therefore, we expect this novel mechanism of CdGM Fermi surface to {\it work for other type-II SCs with a non-DSM normal state.}            

{\it Conclusions and Discussions.---} To summarize, we have introduced the concept of CdGM Fermi surfaces as a new emergent topological response of type-II SCs. Our main finding is that the CdGM Fermi surfaces necessarily arise at zero energy in an Abrikosov lattice when each vortex line is nodal. These in-gap Fermi surfaces are topologically protected for carrying a $\mathbb{Z}_2$ Pfaffian index, which is a natural outcome of an effective gauge structure induced by both the SC phase winding and the physical magnetic field in an ordered vortex lattice. We have provided both an effective model description and a microscopic lattice model simulation for the CdGM Fermi surfaces. Remarkably, this exotic phase is found to persist even when the nodal vortex condition is relaxed. 

Nodal vortices have been predicted to exist in SCing DSMs such as LiFeAs~\cite{Zhang19}, Sr$_{3-x}$SnO~\cite{oudah2016dirac}, Au$_2$Pb~\cite{Schoop15}, etc., as well as in half-Heusler superconductors~\cite{hu2023topological} such as YPtBi~\cite{kim2018beyond} and LuPdBi~\cite{ishihara2021heusler}. Following our theory, the CdGM Fermi surfaces should show up in these systems once the applied ${\bf B}$ exceeds the corresponding lower critical field $H_{c1}$. In particular, a highly ordered vortex lattice has been experimentally observed in LiFeAs by multiple groups~\cite{Zhang19b,Hoffmann22,li2022ordered}, which offers an ideal real-world platform to detect the proposed topological metallicity. The presence of a CdGM Fermi surface indicates a large density of states at $E=0$, which only appear when ${\bf B}$ is turned on. We thus expect the emergent Fermi surface to respond to various experimental probes including scanning tunneling microscopy, tunneling spectroscopy, specific heat measurement, etc~\cite{lapp2020BdG}. Therefore, we are confident that signatures of the CdGM Fermi surfaces will soon be observed in experiments. 

{\it Acknowledgement -} We thank Y. Wang, K. Sun, D. F. Agterberg, C.-K. Chiu, L. Kong, F. Yang, and especially L.-H. Hu for helpful discussions. S.P. and R.-X.Z. are supported by a start-up fund at the University of Tennessee. J.S. acknowledges support from the Joint Quantum Institute and support by the Laboratory for Physical Sciences through its continuous support of the Condensed Matter Theory Center at the University of Maryland.

\bibliography{biblio_TSC}

%apsrev4-2.bst 2019-01-14 (MD) hand-edited version of apsrev4-1.bst
%Control: key (0)
%Control: author (8) initials jnrlst
%Control: editor formatted (1) identically to author
%Control: production of article title (0) allowed
%Control: page (0) single
%Control: year (1) truncated
%Control: production of eprint (0) enabled
\begin{thebibliography}{48}%
\makeatletter
\providecommand \@ifxundefined [1]{%
 \@ifx{#1\undefined}
}%
\providecommand \@ifnum [1]{%
 \ifnum #1\expandafter \@firstoftwo
 \else \expandafter \@secondoftwo
 \fi
}%
\providecommand \@ifx [1]{%
 \ifx #1\expandafter \@firstoftwo
 \else \expandafter \@secondoftwo
 \fi
}%
\providecommand \natexlab [1]{#1}%
\providecommand \enquote  [1]{``#1''}%
\providecommand \bibnamefont  [1]{#1}%
\providecommand \bibfnamefont [1]{#1}%
\providecommand \citenamefont [1]{#1}%
\providecommand \href@noop [0]{\@secondoftwo}%
\providecommand \href [0]{\begingroup \@sanitize@url \@href}%
\providecommand \@href[1]{\@@startlink{#1}\@@href}%
\providecommand \@@href[1]{\endgroup#1\@@endlink}%
\providecommand \@sanitize@url [0]{\catcode `\\12\catcode `\$12\catcode
  `\&12\catcode `\#12\catcode `\^12\catcode `\_12\catcode `\%12\relax}%
\providecommand \@@startlink[1]{}%
\providecommand \@@endlink[0]{}%
\providecommand \url  [0]{\begingroup\@sanitize@url \@url }%
\providecommand \@url [1]{\endgroup\@href {#1}{\urlprefix }}%
\providecommand \urlprefix  [0]{URL }%
\providecommand \Eprint [0]{\href }%
\providecommand \doibase [0]{https://doi.org/}%
\providecommand \selectlanguage [0]{\@gobble}%
\providecommand \bibinfo  [0]{\@secondoftwo}%
\providecommand \bibfield  [0]{\@secondoftwo}%
\providecommand \translation [1]{[#1]}%
\providecommand \BibitemOpen [0]{}%
\providecommand \bibitemStop [0]{}%
\providecommand \bibitemNoStop [0]{.\EOS\space}%
\providecommand \EOS [0]{\spacefactor3000\relax}%
\providecommand \BibitemShut  [1]{\csname bibitem#1\endcsname}%
\let\auto@bib@innerbib\@empty
%</preamble>
\bibitem [{\citenamefont {Bardeen}\ \emph {et~al.}(1957)\citenamefont
  {Bardeen}, \citenamefont {Cooper},\ and\ \citenamefont
  {Schrieffer}}]{BCS1957}%
  \BibitemOpen
  \bibfield  {author} {\bibinfo {author} {\bibfnamefont {J.}~\bibnamefont
  {Bardeen}}, \bibinfo {author} {\bibfnamefont {L.~N.}\ \bibnamefont
  {Cooper}},\ and\ \bibinfo {author} {\bibfnamefont {J.~R.}\ \bibnamefont
  {Schrieffer}},\ }\bibfield  {title} {\bibinfo {title} {Microscopic theory of
  superconductivity},\ }\href {https://doi.org/10.1103/PhysRev.106.162}
  {\bibfield  {journal} {\bibinfo  {journal} {Phys. Rev.}\ }\textbf {\bibinfo
  {volume} {106}},\ \bibinfo {pages} {162} (\bibinfo {year}
  {1957})}\BibitemShut {NoStop}%
\bibitem [{\citenamefont {Gr\"uner}(1988)}]{gruner1988cdw}%
  \BibitemOpen
  \bibfield  {author} {\bibinfo {author} {\bibfnamefont {G.}~\bibnamefont
  {Gr\"uner}},\ }\bibfield  {title} {\bibinfo {title} {The dynamics of
  charge-density waves},\ }\href {https://doi.org/10.1103/RevModPhys.60.1129}
  {\bibfield  {journal} {\bibinfo  {journal} {Rev. Mod. Phys.}\ }\textbf
  {\bibinfo {volume} {60}},\ \bibinfo {pages} {1129} (\bibinfo {year}
  {1988})}\BibitemShut {NoStop}%
\bibitem [{\citenamefont {Bednorz}\ and\ \citenamefont
  {M{\"u}ller}(1986)}]{bednorz1986possible}%
  \BibitemOpen
  \bibfield  {author} {\bibinfo {author} {\bibfnamefont {J.~G.}\ \bibnamefont
  {Bednorz}}\ and\ \bibinfo {author} {\bibfnamefont {K.~A.}\ \bibnamefont
  {M{\"u}ller}},\ }\bibfield  {title} {\bibinfo {title} {Possible high t c
  superconductivity in the ba- la- cu- o system},\ }\href
  {https://doi.org/10.1007/BF01303701} {\bibfield  {journal} {\bibinfo
  {journal} {Zeitschrift f{\"u}r Physik B Condensed Matter}\ }\textbf {\bibinfo
  {volume} {64}},\ \bibinfo {pages} {189} (\bibinfo {year} {1986})}\BibitemShut
  {NoStop}%
\bibitem [{\citenamefont {Sigrist}\ and\ \citenamefont
  {Ueda}(1991)}]{sigrist1991rmp}%
  \BibitemOpen
  \bibfield  {author} {\bibinfo {author} {\bibfnamefont {M.}~\bibnamefont
  {Sigrist}}\ and\ \bibinfo {author} {\bibfnamefont {K.}~\bibnamefont {Ueda}},\
  }\bibfield  {title} {\bibinfo {title} {Phenomenological theory of
  unconventional superconductivity},\ }\href
  {https://doi.org/10.1103/RevModPhys.63.239} {\bibfield  {journal} {\bibinfo
  {journal} {Rev. Mod. Phys.}\ }\textbf {\bibinfo {volume} {63}},\ \bibinfo
  {pages} {239} (\bibinfo {year} {1991})}\BibitemShut {NoStop}%
\bibitem [{\citenamefont {Tsuei}\ and\ \citenamefont
  {Kirtley}(2000)}]{tsuei2000cuprate}%
  \BibitemOpen
  \bibfield  {author} {\bibinfo {author} {\bibfnamefont {C.~C.}\ \bibnamefont
  {Tsuei}}\ and\ \bibinfo {author} {\bibfnamefont {J.~R.}\ \bibnamefont
  {Kirtley}},\ }\bibfield  {title} {\bibinfo {title} {Pairing symmetry in
  cuprate superconductors},\ }\href {https://doi.org/10.1103/RevModPhys.72.969}
  {\bibfield  {journal} {\bibinfo  {journal} {Rev. Mod. Phys.}\ }\textbf
  {\bibinfo {volume} {72}},\ \bibinfo {pages} {969} (\bibinfo {year}
  {2000})}\BibitemShut {NoStop}%
\bibitem [{\citenamefont {Hirschfeld}\ \emph {et~al.}(2011)\citenamefont
  {Hirschfeld}, \citenamefont {Korshunov},\ and\ \citenamefont
  {Mazin}}]{hirschfeld2011gap}%
  \BibitemOpen
  \bibfield  {author} {\bibinfo {author} {\bibfnamefont {P.}~\bibnamefont
  {Hirschfeld}}, \bibinfo {author} {\bibfnamefont {M.}~\bibnamefont
  {Korshunov}},\ and\ \bibinfo {author} {\bibfnamefont {I.}~\bibnamefont
  {Mazin}},\ }\bibfield  {title} {\bibinfo {title} {Gap symmetry and structure
  of fe-based superconductors},\ }\href
  {https://iopscience.iop.org/article/10.1088/0034-4885/74/12/124508}
  {\bibfield  {journal} {\bibinfo  {journal} {Reports on Progress in Physics}\
  }\textbf {\bibinfo {volume} {74}},\ \bibinfo {pages} {124508} (\bibinfo
  {year} {2011})}\BibitemShut {NoStop}%
\bibitem [{\citenamefont {Zhao}\ and\ \citenamefont
  {Wang}(2013)}]{zhao2013topo}%
  \BibitemOpen
  \bibfield  {author} {\bibinfo {author} {\bibfnamefont {Y.~X.}\ \bibnamefont
  {Zhao}}\ and\ \bibinfo {author} {\bibfnamefont {Z.~D.}\ \bibnamefont
  {Wang}},\ }\bibfield  {title} {\bibinfo {title} {Topological classification
  and stability of fermi surfaces},\ }\href
  {https://doi.org/10.1103/PhysRevLett.110.240404} {\bibfield  {journal}
  {\bibinfo  {journal} {Phys. Rev. Lett.}\ }\textbf {\bibinfo {volume} {110}},\
  \bibinfo {pages} {240404} (\bibinfo {year} {2013})}\BibitemShut {NoStop}%
\bibitem [{\citenamefont {Kobayashi}\ \emph {et~al.}(2014)\citenamefont
  {Kobayashi}, \citenamefont {Shiozaki}, \citenamefont {Tanaka},\ and\
  \citenamefont {Sato}}]{kobayashi2014blount}%
  \BibitemOpen
  \bibfield  {author} {\bibinfo {author} {\bibfnamefont {S.}~\bibnamefont
  {Kobayashi}}, \bibinfo {author} {\bibfnamefont {K.}~\bibnamefont {Shiozaki}},
  \bibinfo {author} {\bibfnamefont {Y.}~\bibnamefont {Tanaka}},\ and\ \bibinfo
  {author} {\bibfnamefont {M.}~\bibnamefont {Sato}},\ }\bibfield  {title}
  {\bibinfo {title} {Topological blount's theorem of odd-parity
  superconductors},\ }\href {https://doi.org/10.1103/PhysRevB.90.024516}
  {\bibfield  {journal} {\bibinfo  {journal} {Phys. Rev. B}\ }\textbf {\bibinfo
  {volume} {90}},\ \bibinfo {pages} {024516} (\bibinfo {year}
  {2014})}\BibitemShut {NoStop}%
\bibitem [{\citenamefont {Zhao}\ \emph {et~al.}(2016)\citenamefont {Zhao},
  \citenamefont {Schnyder},\ and\ \citenamefont {Wang}}]{zhao2016unified}%
  \BibitemOpen
  \bibfield  {author} {\bibinfo {author} {\bibfnamefont {Y.~X.}\ \bibnamefont
  {Zhao}}, \bibinfo {author} {\bibfnamefont {A.~P.}\ \bibnamefont {Schnyder}},\
  and\ \bibinfo {author} {\bibfnamefont {Z.~D.}\ \bibnamefont {Wang}},\
  }\bibfield  {title} {\bibinfo {title} {Unified theory of $pt$ and $cp$
  invariant topological metals and nodal superconductors},\ }\href
  {https://doi.org/10.1103/PhysRevLett.116.156402} {\bibfield  {journal}
  {\bibinfo  {journal} {Phys. Rev. Lett.}\ }\textbf {\bibinfo {volume} {116}},\
  \bibinfo {pages} {156402} (\bibinfo {year} {2016})}\BibitemShut {NoStop}%
\bibitem [{\citenamefont {Agterberg}\ \emph {et~al.}(2017)\citenamefont
  {Agterberg}, \citenamefont {Brydon},\ and\ \citenamefont
  {Timm}}]{Agterberg17}%
  \BibitemOpen
  \bibfield  {author} {\bibinfo {author} {\bibfnamefont {D.~F.}\ \bibnamefont
  {Agterberg}}, \bibinfo {author} {\bibfnamefont {P.~M.~R.}\ \bibnamefont
  {Brydon}},\ and\ \bibinfo {author} {\bibfnamefont {C.}~\bibnamefont {Timm}},\
  }\bibfield  {title} {\bibinfo {title} {Bogoliubov fermi surfaces in
  superconductors with broken time-reversal symmetry},\ }\href
  {https://doi.org/10.1103/PhysRevLett.118.127001} {\bibfield  {journal}
  {\bibinfo  {journal} {Phys. Rev. Lett.}\ }\textbf {\bibinfo {volume} {118}},\
  \bibinfo {pages} {127001} (\bibinfo {year} {2017})}\BibitemShut {NoStop}%
\bibitem [{\citenamefont {Brydon}\ \emph {et~al.}(2018)\citenamefont {Brydon},
  \citenamefont {Agterberg}, \citenamefont {Menke},\ and\ \citenamefont
  {Timm}}]{brydon2018BdG}%
  \BibitemOpen
  \bibfield  {author} {\bibinfo {author} {\bibfnamefont {P.~M.~R.}\
  \bibnamefont {Brydon}}, \bibinfo {author} {\bibfnamefont {D.~F.}\
  \bibnamefont {Agterberg}}, \bibinfo {author} {\bibfnamefont {H.}~\bibnamefont
  {Menke}},\ and\ \bibinfo {author} {\bibfnamefont {C.}~\bibnamefont {Timm}},\
  }\bibfield  {title} {\bibinfo {title} {Bogoliubov fermi surfaces: General
  theory, magnetic order, and topology},\ }\href
  {https://doi.org/10.1103/PhysRevB.98.224509} {\bibfield  {journal} {\bibinfo
  {journal} {Phys. Rev. B}\ }\textbf {\bibinfo {volume} {98}},\ \bibinfo
  {pages} {224509} (\bibinfo {year} {2018})}\BibitemShut {NoStop}%
\bibitem [{\citenamefont {Santos}\ \emph {et~al.}(2019)\citenamefont {Santos},
  \citenamefont {Wang},\ and\ \citenamefont {Fradkin}}]{santos2019pdw}%
  \BibitemOpen
  \bibfield  {author} {\bibinfo {author} {\bibfnamefont {L.~H.}\ \bibnamefont
  {Santos}}, \bibinfo {author} {\bibfnamefont {Y.}~\bibnamefont {Wang}},\ and\
  \bibinfo {author} {\bibfnamefont {E.}~\bibnamefont {Fradkin}},\ }\bibfield
  {title} {\bibinfo {title} {Pair-density-wave order and paired fractional
  quantum hall fluids},\ }\href {https://doi.org/10.1103/PhysRevX.9.021047}
  {\bibfield  {journal} {\bibinfo  {journal} {Phys. Rev. X}\ }\textbf {\bibinfo
  {volume} {9}},\ \bibinfo {pages} {021047} (\bibinfo {year}
  {2019})}\BibitemShut {NoStop}%
\bibitem [{\citenamefont {Link}\ and\ \citenamefont
  {Herbut}(2020)}]{link2020BdG}%
  \BibitemOpen
  \bibfield  {author} {\bibinfo {author} {\bibfnamefont {J.~M.}\ \bibnamefont
  {Link}}\ and\ \bibinfo {author} {\bibfnamefont {I.~F.}\ \bibnamefont
  {Herbut}},\ }\bibfield  {title} {\bibinfo {title} {Bogoliubov-fermi surfaces
  in noncentrosymmetric multicomponent superconductors},\ }\href
  {https://doi.org/10.1103/PhysRevLett.125.237004} {\bibfield  {journal}
  {\bibinfo  {journal} {Phys. Rev. Lett.}\ }\textbf {\bibinfo {volume} {125}},\
  \bibinfo {pages} {237004} (\bibinfo {year} {2020})}\BibitemShut {NoStop}%
\bibitem [{\citenamefont {Shaffer}\ \emph {et~al.}(2020)\citenamefont
  {Shaffer}, \citenamefont {Kang}, \citenamefont {Burnell},\ and\ \citenamefont
  {Fernandes}}]{shaffer2020crystalline}%
  \BibitemOpen
  \bibfield  {author} {\bibinfo {author} {\bibfnamefont {D.}~\bibnamefont
  {Shaffer}}, \bibinfo {author} {\bibfnamefont {J.}~\bibnamefont {Kang}},
  \bibinfo {author} {\bibfnamefont {F.~J.}\ \bibnamefont {Burnell}},\ and\
  \bibinfo {author} {\bibfnamefont {R.~M.}\ \bibnamefont {Fernandes}},\
  }\bibfield  {title} {\bibinfo {title} {Crystalline nodal topological
  superconductivity and bogolyubov fermi surfaces in monolayer
  ${\mathrm{nbse}}_{2}$},\ }\href {https://doi.org/10.1103/PhysRevB.101.224503}
  {\bibfield  {journal} {\bibinfo  {journal} {Phys. Rev. B}\ }\textbf {\bibinfo
  {volume} {101}},\ \bibinfo {pages} {224503} (\bibinfo {year}
  {2020})}\BibitemShut {NoStop}%
\bibitem [{\citenamefont {Setty}\ \emph {et~al.}(2020)\citenamefont {Setty},
  \citenamefont {Cao}, \citenamefont {Kreisel}, \citenamefont {Bhattacharyya},\
  and\ \citenamefont {Hirschfeld}}]{setty2020BdG}%
  \BibitemOpen
  \bibfield  {author} {\bibinfo {author} {\bibfnamefont {C.}~\bibnamefont
  {Setty}}, \bibinfo {author} {\bibfnamefont {Y.}~\bibnamefont {Cao}}, \bibinfo
  {author} {\bibfnamefont {A.}~\bibnamefont {Kreisel}}, \bibinfo {author}
  {\bibfnamefont {S.}~\bibnamefont {Bhattacharyya}},\ and\ \bibinfo {author}
  {\bibfnamefont {P.~J.}\ \bibnamefont {Hirschfeld}},\ }\bibfield  {title}
  {\bibinfo {title} {Bogoliubov fermi surfaces in spin-$\frac{1}{2}$ systems:
  Model hamiltonians and experimental consequences},\ }\href
  {https://doi.org/10.1103/PhysRevB.102.064504} {\bibfield  {journal} {\bibinfo
   {journal} {Phys. Rev. B}\ }\textbf {\bibinfo {volume} {102}},\ \bibinfo
  {pages} {064504} (\bibinfo {year} {2020})}\BibitemShut {NoStop}%
\bibitem [{\citenamefont {Dutta}\ \emph {et~al.}(2021)\citenamefont {Dutta},
  \citenamefont {Parhizgar},\ and\ \citenamefont
  {Black-Schaffer}}]{dutta2021bdg}%
  \BibitemOpen
  \bibfield  {author} {\bibinfo {author} {\bibfnamefont {P.}~\bibnamefont
  {Dutta}}, \bibinfo {author} {\bibfnamefont {F.}~\bibnamefont {Parhizgar}},\
  and\ \bibinfo {author} {\bibfnamefont {A.~M.}\ \bibnamefont
  {Black-Schaffer}},\ }\bibfield  {title} {\bibinfo {title} {Superconductivity
  in spin-$3/2$ systems: Symmetry classification, odd-frequency pairs, and
  bogoliubov fermi surfaces},\ }\href
  {https://doi.org/10.1103/PhysRevResearch.3.033255} {\bibfield  {journal}
  {\bibinfo  {journal} {Phys. Rev. Res.}\ }\textbf {\bibinfo {volume} {3}},\
  \bibinfo {pages} {033255} (\bibinfo {year} {2021})}\BibitemShut {NoStop}%
\bibitem [{\citenamefont {Volkov}\ \emph {et~al.}(2023)\citenamefont {Volkov},
  \citenamefont {Wilson}, \citenamefont {Lucht},\ and\ \citenamefont
  {Pixley}}]{volkov2023twisted}%
  \BibitemOpen
  \bibfield  {author} {\bibinfo {author} {\bibfnamefont {P.~A.}\ \bibnamefont
  {Volkov}}, \bibinfo {author} {\bibfnamefont {J.~H.}\ \bibnamefont {Wilson}},
  \bibinfo {author} {\bibfnamefont {K.~P.}\ \bibnamefont {Lucht}},\ and\
  \bibinfo {author} {\bibfnamefont {J.~H.}\ \bibnamefont {Pixley}},\ }\bibfield
   {title} {\bibinfo {title} {Current- and field-induced topology in twisted
  nodal superconductors},\ }\href
  {https://doi.org/10.1103/PhysRevLett.130.186001} {\bibfield  {journal}
  {\bibinfo  {journal} {Phys. Rev. Lett.}\ }\textbf {\bibinfo {volume} {130}},\
  \bibinfo {pages} {186001} (\bibinfo {year} {2023})}\BibitemShut {NoStop}%
\bibitem [{\citenamefont {Tsubaki~Nagashima}(2022)}]{Nagashima22}%
  \BibitemOpen
  \bibfield  {author} {\bibinfo {author} {\bibfnamefont {S.~N. e.~a.}\
  \bibnamefont {Tsubaki~Nagashima}, \bibfnamefont {Takahiro~Hashimoto}},\
  }\href {https://doi.org/10.21203/rs.3.rs-2224728/v1} {\bibinfo {title}
  {Discovery of nematic bogoliubov fermi surface in an iron-chalcogenide
  superconductor}} (\bibinfo {year} {2022})\BibitemShut {NoStop}%
\bibitem [{\citenamefont {Mizukami}\ \emph {et~al.}(2023)\citenamefont
  {Mizukami}, \citenamefont {Haze}, \citenamefont {Tanaka}, \citenamefont
  {Matsuura}, \citenamefont {Sano}, \citenamefont {Böker}, \citenamefont
  {Eremin}, \citenamefont {Kasahara}, \citenamefont {Matsuda},\ and\
  \citenamefont {Shibauchi}}]{Mizukami23}%
  \BibitemOpen
  \bibfield  {author} {\bibinfo {author} {\bibfnamefont {Y.}~\bibnamefont
  {Mizukami}}, \bibinfo {author} {\bibfnamefont {M.}~\bibnamefont {Haze}},
  \bibinfo {author} {\bibfnamefont {O.}~\bibnamefont {Tanaka}}, \bibinfo
  {author} {\bibfnamefont {K.}~\bibnamefont {Matsuura}}, \bibinfo {author}
  {\bibfnamefont {D.}~\bibnamefont {Sano}}, \bibinfo {author} {\bibfnamefont
  {J.}~\bibnamefont {Böker}}, \bibinfo {author} {\bibfnamefont
  {I.}~\bibnamefont {Eremin}}, \bibinfo {author} {\bibfnamefont
  {S.}~\bibnamefont {Kasahara}}, \bibinfo {author} {\bibfnamefont
  {Y.}~\bibnamefont {Matsuda}},\ and\ \bibinfo {author} {\bibfnamefont
  {T.}~\bibnamefont {Shibauchi}},\ }\bibfield  {title} {\bibinfo {title}
  {Unusual crossover from bardeen-cooper-schrieffer to bose-einstein-condensate
  superconductivity in iron chalcogenides},\ }\bibfield  {journal} {\bibinfo
  {journal} {Communications Physics}\ }\textbf {\bibinfo {volume} {6}},\ \href
  {https://doi.org/10.1038/s42005-023-01289-8} {10.1038/s42005-023-01289-8}
  (\bibinfo {year} {2023})\BibitemShut {NoStop}%
\bibitem [{\citenamefont {Matsuura}\ \emph {et~al.}(2023)\citenamefont
  {Matsuura}, \citenamefont {Roppongi}, \citenamefont {Qiu}, \citenamefont
  {Sheng}, \citenamefont {Cai}, \citenamefont {Yamakawa}, \citenamefont
  {Guguchia}, \citenamefont {Day}, \citenamefont {Kojima}, \citenamefont
  {Damascelli} \emph {et~al.}}]{Matsuura23}%
  \BibitemOpen
  \bibfield  {author} {\bibinfo {author} {\bibfnamefont {K.}~\bibnamefont
  {Matsuura}}, \bibinfo {author} {\bibfnamefont {M.}~\bibnamefont {Roppongi}},
  \bibinfo {author} {\bibfnamefont {M.}~\bibnamefont {Qiu}}, \bibinfo {author}
  {\bibfnamefont {Q.}~\bibnamefont {Sheng}}, \bibinfo {author} {\bibfnamefont
  {Y.}~\bibnamefont {Cai}}, \bibinfo {author} {\bibfnamefont {K.}~\bibnamefont
  {Yamakawa}}, \bibinfo {author} {\bibfnamefont {Z.}~\bibnamefont {Guguchia}},
  \bibinfo {author} {\bibfnamefont {R.~P.}\ \bibnamefont {Day}}, \bibinfo
  {author} {\bibfnamefont {K.~M.}\ \bibnamefont {Kojima}}, \bibinfo {author}
  {\bibfnamefont {A.}~\bibnamefont {Damascelli}}, \emph {et~al.},\ }\bibfield
  {title} {\bibinfo {title} {Two superconducting states with broken
  time-reversal symmetry in fese1- x s x},\ }\href
  {https://www.pnas.org/doi/abs/10.1073/pnas.2208276120} {\bibfield  {journal}
  {\bibinfo  {journal} {Proceedings of the National Academy of Sciences}\
  }\textbf {\bibinfo {volume} {120}},\ \bibinfo {pages} {e2208276120} (\bibinfo
  {year} {2023})}\BibitemShut {NoStop}%
\bibitem [{\citenamefont {Wu}\ \emph {et~al.}(2023)\citenamefont {Wu},
  \citenamefont {Amin}, \citenamefont {Yu},\ and\ \citenamefont
  {Agterberg}}]{Wu23}%
  \BibitemOpen
  \bibfield  {author} {\bibinfo {author} {\bibfnamefont {H.}~\bibnamefont
  {Wu}}, \bibinfo {author} {\bibfnamefont {A.}~\bibnamefont {Amin}}, \bibinfo
  {author} {\bibfnamefont {Y.}~\bibnamefont {Yu}},\ and\ \bibinfo {author}
  {\bibfnamefont {D.~F.}\ \bibnamefont {Agterberg}},\ }\href
  {https://doi.org/10.48550/arXiv.2306.11200} {\bibinfo {title} {Nematic
  bogoliubov fermi surfaces from magnetic toroidal order in
  fe{Se}$_{1-x}${S}$_x$}} (\bibinfo {year} {2023}),\ \Eprint
  {https://arxiv.org/abs/2306.11200} {arXiv:2306.11200 [cond-mat.supr-con]}
  \BibitemShut {NoStop}%
\bibitem [{\citenamefont {Abrikosov}(1957)}]{Abrikosov56}%
  \BibitemOpen
  \bibfield  {author} {\bibinfo {author} {\bibfnamefont {A.~A.}\ \bibnamefont
  {Abrikosov}},\ }\bibfield  {title} {\bibinfo {title} {{On the Magnetic
  properties of superconductors of the second group}},\ }\href@noop {}
  {\bibfield  {journal} {\bibinfo  {journal} {Sov. Phys. JETP}\ }\textbf
  {\bibinfo {volume} {5}},\ \bibinfo {pages} {1174} (\bibinfo {year}
  {1957})}\BibitemShut {NoStop}%
\bibitem [{\citenamefont {Franz}\ and\ \citenamefont {Te\ifmmode \check{s}\else
  \v{s}\fi{}anovi\ifmmode~\acute{c}\else \'{c}\fi{}}(2000)}]{Franz00}%
  \BibitemOpen
  \bibfield  {author} {\bibinfo {author} {\bibfnamefont {M.}~\bibnamefont
  {Franz}}\ and\ \bibinfo {author} {\bibfnamefont {Z.}~\bibnamefont {Te\ifmmode
  \check{s}\else \v{s}\fi{}anovi\ifmmode~\acute{c}\else \'{c}\fi{}}},\
  }\bibfield  {title} {\bibinfo {title} {Quasiparticles in the vortex lattice
  of unconventional superconductors: Bloch waves or landau levels?},\ }\href
  {https://doi.org/10.1103/PhysRevLett.84.554} {\bibfield  {journal} {\bibinfo
  {journal} {Phys. Rev. Lett.}\ }\textbf {\bibinfo {volume} {84}},\ \bibinfo
  {pages} {554} (\bibinfo {year} {2000})}\BibitemShut {NoStop}%
\bibitem [{\citenamefont {Vafek}\ \emph {et~al.}(2001)\citenamefont {Vafek},
  \citenamefont {Melikyan}, \citenamefont {Franz},\ and\ \citenamefont
  {Te\ifmmode \check{s}\else \v{s}\fi{}anovi\ifmmode~\acute{c}\else
  \'{c}\fi{}}}]{Vafek01}%
  \BibitemOpen
  \bibfield  {author} {\bibinfo {author} {\bibfnamefont {O.}~\bibnamefont
  {Vafek}}, \bibinfo {author} {\bibfnamefont {A.}~\bibnamefont {Melikyan}},
  \bibinfo {author} {\bibfnamefont {M.}~\bibnamefont {Franz}},\ and\ \bibinfo
  {author} {\bibfnamefont {Z.}~\bibnamefont {Te\ifmmode \check{s}\else
  \v{s}\fi{}anovi\ifmmode~\acute{c}\else \'{c}\fi{}}},\ }\bibfield  {title}
  {\bibinfo {title} {Quasiparticles and vortices in unconventional
  superconductors},\ }\href {https://doi.org/10.1103/PhysRevB.63.134509}
  {\bibfield  {journal} {\bibinfo  {journal} {Phys. Rev. B}\ }\textbf {\bibinfo
  {volume} {63}},\ \bibinfo {pages} {134509} (\bibinfo {year}
  {2001})}\BibitemShut {NoStop}%
\bibitem [{\citenamefont {Chiu}\ \emph {et~al.}(2015)\citenamefont {Chiu},
  \citenamefont {Pikulin},\ and\ \citenamefont {Franz}}]{chiu2015strongly}%
  \BibitemOpen
  \bibfield  {author} {\bibinfo {author} {\bibfnamefont {C.-K.}\ \bibnamefont
  {Chiu}}, \bibinfo {author} {\bibfnamefont {D.~I.}\ \bibnamefont {Pikulin}},\
  and\ \bibinfo {author} {\bibfnamefont {M.}~\bibnamefont {Franz}},\ }\bibfield
   {title} {\bibinfo {title} {Strongly interacting majorana fermions},\ }\href
  {https://doi.org/10.1103/PhysRevB.91.165402} {\bibfield  {journal} {\bibinfo
  {journal} {Phys. Rev. B}\ }\textbf {\bibinfo {volume} {91}},\ \bibinfo
  {pages} {165402} (\bibinfo {year} {2015})}\BibitemShut {NoStop}%
\bibitem [{\citenamefont {Liu}\ and\ \citenamefont {Franz}(2015)}]{Liu15}%
  \BibitemOpen
  \bibfield  {author} {\bibinfo {author} {\bibfnamefont {T.}~\bibnamefont
  {Liu}}\ and\ \bibinfo {author} {\bibfnamefont {M.}~\bibnamefont {Franz}},\
  }\bibfield  {title} {\bibinfo {title} {Electronic structure of topological
  superconductors in the presence of a vortex lattice},\ }\href
  {https://doi.org/10.1103/PhysRevB.92.134519} {\bibfield  {journal} {\bibinfo
  {journal} {Phys. Rev. B}\ }\textbf {\bibinfo {volume} {92}},\ \bibinfo
  {pages} {134519} (\bibinfo {year} {2015})}\BibitemShut {NoStop}%
\bibitem [{\citenamefont {Chaudhary}\ and\ \citenamefont
  {MacDonald}(2020)}]{chaudhary2020vortex}%
  \BibitemOpen
  \bibfield  {author} {\bibinfo {author} {\bibfnamefont {G.}~\bibnamefont
  {Chaudhary}}\ and\ \bibinfo {author} {\bibfnamefont {A.~H.}\ \bibnamefont
  {MacDonald}},\ }\bibfield  {title} {\bibinfo {title} {Vortex-lattice
  structure and topological superconductivity in the quantum hall regime},\
  }\href {https://doi.org/10.1103/PhysRevB.101.024516} {\bibfield  {journal}
  {\bibinfo  {journal} {Phys. Rev. B}\ }\textbf {\bibinfo {volume} {101}},\
  \bibinfo {pages} {024516} (\bibinfo {year} {2020})}\BibitemShut {NoStop}%
\bibitem [{\citenamefont {Schirmer}\ \emph {et~al.}(2022)\citenamefont
  {Schirmer}, \citenamefont {Liu},\ and\ \citenamefont
  {Jain}}]{schirmer2022phase}%
  \BibitemOpen
  \bibfield  {author} {\bibinfo {author} {\bibfnamefont {J.}~\bibnamefont
  {Schirmer}}, \bibinfo {author} {\bibfnamefont {C.-X.}\ \bibnamefont {Liu}},\
  and\ \bibinfo {author} {\bibfnamefont {J.}~\bibnamefont {Jain}},\ }\bibfield
  {title} {\bibinfo {title} {Phase diagram of superconductivity in the integer
  quantum hall regime},\ }\href {https://doi.org/10.1073/pnas.220294811}
  {\bibfield  {journal} {\bibinfo  {journal} {Proceedings of the National
  Academy of Sciences}\ }\textbf {\bibinfo {volume} {119}},\ \bibinfo {pages}
  {e2202948119} (\bibinfo {year} {2022})}\BibitemShut {NoStop}%
\bibitem [{\citenamefont {Caroli}\ \emph {et~al.}(1964)\citenamefont {Caroli},
  \citenamefont {{De Gennes}},\ and\ \citenamefont {Matricon}}]{Caroli64}%
  \BibitemOpen
  \bibfield  {author} {\bibinfo {author} {\bibfnamefont {C.}~\bibnamefont
  {Caroli}}, \bibinfo {author} {\bibfnamefont {P.}~\bibnamefont {{De
  Gennes}}},\ and\ \bibinfo {author} {\bibfnamefont {J.}~\bibnamefont
  {Matricon}},\ }\bibfield  {title} {\bibinfo {title} {Bound fermion states on
  a vortex line in a type ii superconductor},\ }\href
  {https://doi.org/https://doi.org/10.1016/0031-9163(64)90375-0} {\bibfield
  {journal} {\bibinfo  {journal} {Physics Letters}\ }\textbf {\bibinfo {volume}
  {9}},\ \bibinfo {pages} {307} (\bibinfo {year} {1964})}\BibitemShut {NoStop}%
\bibitem [{\citenamefont {Qin}\ \emph {et~al.}(2019)\citenamefont {Qin},
  \citenamefont {Hu}, \citenamefont {Le}, \citenamefont {Zeng}, \citenamefont
  {Zhang}, \citenamefont {Fang},\ and\ \citenamefont {Hu}}]{Qin19}%
  \BibitemOpen
  \bibfield  {author} {\bibinfo {author} {\bibfnamefont {S.}~\bibnamefont
  {Qin}}, \bibinfo {author} {\bibfnamefont {L.}~\bibnamefont {Hu}}, \bibinfo
  {author} {\bibfnamefont {C.}~\bibnamefont {Le}}, \bibinfo {author}
  {\bibfnamefont {J.}~\bibnamefont {Zeng}}, \bibinfo {author} {\bibfnamefont
  {F.-c.}\ \bibnamefont {Zhang}}, \bibinfo {author} {\bibfnamefont
  {C.}~\bibnamefont {Fang}},\ and\ \bibinfo {author} {\bibfnamefont
  {J.}~\bibnamefont {Hu}},\ }\bibfield  {title} {\bibinfo {title} {Quasi-1d
  topological nodal vortex line phase in doped superconducting 3d dirac
  semimetals},\ }\href {https://doi.org/10.1103/PhysRevLett.123.027003}
  {\bibfield  {journal} {\bibinfo  {journal} {Phys. Rev. Lett.}\ }\textbf
  {\bibinfo {volume} {123}},\ \bibinfo {pages} {027003} (\bibinfo {year}
  {2019})}\BibitemShut {NoStop}%
\bibitem [{\citenamefont {K\"onig}\ and\ \citenamefont
  {Coleman}(2019)}]{Konig19}%
  \BibitemOpen
  \bibfield  {author} {\bibinfo {author} {\bibfnamefont {E.~J.}\ \bibnamefont
  {K\"onig}}\ and\ \bibinfo {author} {\bibfnamefont {P.}~\bibnamefont
  {Coleman}},\ }\bibfield  {title} {\bibinfo {title}
  {Crystalline-symmetry-protected helical majorana modes in the iron
  pnictides},\ }\href {https://doi.org/10.1103/PhysRevLett.122.207001}
  {\bibfield  {journal} {\bibinfo  {journal} {Phys. Rev. Lett.}\ }\textbf
  {\bibinfo {volume} {122}},\ \bibinfo {pages} {207001} (\bibinfo {year}
  {2019})}\BibitemShut {NoStop}%
\bibitem [{\citenamefont {Yan}\ \emph {et~al.}(2020)\citenamefont {Yan},
  \citenamefont {Wu},\ and\ \citenamefont {Huang}}]{yan2020vortex}%
  \BibitemOpen
  \bibfield  {author} {\bibinfo {author} {\bibfnamefont {Z.}~\bibnamefont
  {Yan}}, \bibinfo {author} {\bibfnamefont {Z.}~\bibnamefont {Wu}},\ and\
  \bibinfo {author} {\bibfnamefont {W.}~\bibnamefont {Huang}},\ }\bibfield
  {title} {\bibinfo {title} {Vortex end majorana zero modes in superconducting
  dirac and weyl semimetals},\ }\href
  {https://doi.org/10.1103/PhysRevLett.124.257001} {\bibfield  {journal}
  {\bibinfo  {journal} {Phys. Rev. Lett.}\ }\textbf {\bibinfo {volume} {124}},\
  \bibinfo {pages} {257001} (\bibinfo {year} {2020})}\BibitemShut {NoStop}%
\bibitem [{\citenamefont {Hu}\ \emph {et~al.}(2022)\citenamefont {Hu},
  \citenamefont {Wu}, \citenamefont {Liu},\ and\ \citenamefont {Zhang}}]{Hu22}%
  \BibitemOpen
  \bibfield  {author} {\bibinfo {author} {\bibfnamefont {L.-H.}\ \bibnamefont
  {Hu}}, \bibinfo {author} {\bibfnamefont {X.}~\bibnamefont {Wu}}, \bibinfo
  {author} {\bibfnamefont {C.-X.}\ \bibnamefont {Liu}},\ and\ \bibinfo {author}
  {\bibfnamefont {R.-X.}\ \bibnamefont {Zhang}},\ }\bibfield  {title} {\bibinfo
  {title} {Competing vortex topologies in iron-based superconductors},\ }\href
  {https://doi.org/10.1103/PhysRevLett.129.277001} {\bibfield  {journal}
  {\bibinfo  {journal} {Phys. Rev. Lett.}\ }\textbf {\bibinfo {volume} {129}},\
  \bibinfo {pages} {277001} (\bibinfo {year} {2022})}\BibitemShut {NoStop}%
\bibitem [{\citenamefont {Hu}\ and\ \citenamefont
  {Zhang}(2023)}]{hu2023topological}%
  \BibitemOpen
  \bibfield  {author} {\bibinfo {author} {\bibfnamefont {L.-H.}\ \bibnamefont
  {Hu}}\ and\ \bibinfo {author} {\bibfnamefont {R.-X.}\ \bibnamefont {Zhang}},\
  }\bibfield  {title} {\bibinfo {title} {Topological superconducting vortex
  from trivial electronic bands},\ }\href
  {https://doi.org/10.1038/s41467-023-36347-w} {\bibfield  {journal} {\bibinfo
  {journal} {Nature Communications}\ }\textbf {\bibinfo {volume} {14}},\
  \bibinfo {pages} {640} (\bibinfo {year} {2023})}\BibitemShut {NoStop}%
\bibitem [{\citenamefont {Hofstadter}(1976)}]{hofstadter1976}%
  \BibitemOpen
  \bibfield  {author} {\bibinfo {author} {\bibfnamefont {D.~R.}\ \bibnamefont
  {Hofstadter}},\ }\bibfield  {title} {\bibinfo {title} {Energy levels and wave
  functions of bloch electrons in rational and irrational magnetic fields},\
  }\href {https://doi.org/10.1103/PhysRevB.14.2239} {\bibfield  {journal}
  {\bibinfo  {journal} {Phys. Rev. B}\ }\textbf {\bibinfo {volume} {14}},\
  \bibinfo {pages} {2239} (\bibinfo {year} {1976})}\BibitemShut {NoStop}%
\bibitem [{SM()}]{SM}%
  \BibitemOpen
  \href@noop {} {}\bibinfo {note} {See the Supplemental Material for detailed
  information about the analytical solution of single-vortex state, the
  calculation of Peierls phase, the symmetry and topological properites of our
  models, and additional numerical results for different vortex lattice sizes,
  which includes Refs.~\cite{Fu08}.}\BibitemShut {Stop}%
\bibitem [{\citenamefont {Haldane}(1988)}]{Haldane88}%
  \BibitemOpen
  \bibfield  {author} {\bibinfo {author} {\bibfnamefont {F.~D.~M.}\
  \bibnamefont {Haldane}},\ }\bibfield  {title} {\bibinfo {title} {Model for a
  quantum hall effect without landau levels: Condensed-matter realization of
  the "parity anomaly"},\ }\href {https://doi.org/10.1103/PhysRevLett.61.2015}
  {\bibfield  {journal} {\bibinfo  {journal} {Phys. Rev. Lett.}\ }\textbf
  {\bibinfo {volume} {61}},\ \bibinfo {pages} {2015} (\bibinfo {year}
  {1988})}\BibitemShut {NoStop}%
\bibitem [{\citenamefont {Wang}\ \emph {et~al.}(2012)\citenamefont {Wang},
  \citenamefont {Sun}, \citenamefont {Chen}, \citenamefont {Franchini},
  \citenamefont {Xu}, \citenamefont {Weng}, \citenamefont {Dai},\ and\
  \citenamefont {Fang}}]{wang2012dirac}%
  \BibitemOpen
  \bibfield  {author} {\bibinfo {author} {\bibfnamefont {Z.}~\bibnamefont
  {Wang}}, \bibinfo {author} {\bibfnamefont {Y.}~\bibnamefont {Sun}}, \bibinfo
  {author} {\bibfnamefont {X.-Q.}\ \bibnamefont {Chen}}, \bibinfo {author}
  {\bibfnamefont {C.}~\bibnamefont {Franchini}}, \bibinfo {author}
  {\bibfnamefont {G.}~\bibnamefont {Xu}}, \bibinfo {author} {\bibfnamefont
  {H.}~\bibnamefont {Weng}}, \bibinfo {author} {\bibfnamefont {X.}~\bibnamefont
  {Dai}},\ and\ \bibinfo {author} {\bibfnamefont {Z.}~\bibnamefont {Fang}},\
  }\bibfield  {title} {\bibinfo {title} {Dirac semimetal and topological phase
  transitions in ${A}_{3}$bi ($a=\text{Na}$, k, rb)},\ }\href
  {https://doi.org/10.1103/PhysRevB.85.195320} {\bibfield  {journal} {\bibinfo
  {journal} {Phys. Rev. B}\ }\textbf {\bibinfo {volume} {85}},\ \bibinfo
  {pages} {195320} (\bibinfo {year} {2012})}\BibitemShut {NoStop}%
\bibitem [{\citenamefont {{Zhang}}\ \emph {et~al.}(2019)\citenamefont
  {{Zhang}}, \citenamefont {{Wang}}, \citenamefont {{Wu}}, \citenamefont
  {{Yaji}}, \citenamefont {{Ishida}}, \citenamefont {{Kohama}}, \citenamefont
  {{Dai}}, \citenamefont {{Sun}}, \citenamefont {{Bareille}}, \citenamefont
  {{Kuroda}}, \citenamefont {{Kondo}}, \citenamefont {{Okazaki}}, \citenamefont
  {{Kindo}}, \citenamefont {{Wang}}, \citenamefont {{Jin}}, \citenamefont
  {{Hu}}, \citenamefont {{Thomale}}, \citenamefont {{Sumida}}, \citenamefont
  {{Wu}}, \citenamefont {{Miyamoto}}, \citenamefont {{Okuda}}, \citenamefont
  {{Ding}}, \citenamefont {{Gu}}, \citenamefont {{Tamegai}}, \citenamefont
  {{Kawakami}}, \citenamefont {{Sato}},\ and\ \citenamefont
  {{Shin}}}]{Zhang19}%
  \BibitemOpen
  \bibfield  {author} {\bibinfo {author} {\bibfnamefont {P.}~\bibnamefont
  {{Zhang}}}, \bibinfo {author} {\bibfnamefont {Z.}~\bibnamefont {{Wang}}},
  \bibinfo {author} {\bibfnamefont {X.}~\bibnamefont {{Wu}}}, \bibinfo {author}
  {\bibfnamefont {K.}~\bibnamefont {{Yaji}}}, \bibinfo {author} {\bibfnamefont
  {Y.}~\bibnamefont {{Ishida}}}, \bibinfo {author} {\bibfnamefont
  {Y.}~\bibnamefont {{Kohama}}}, \bibinfo {author} {\bibfnamefont
  {G.}~\bibnamefont {{Dai}}}, \bibinfo {author} {\bibfnamefont
  {Y.}~\bibnamefont {{Sun}}}, \bibinfo {author} {\bibfnamefont
  {C.}~\bibnamefont {{Bareille}}}, \bibinfo {author} {\bibfnamefont
  {K.}~\bibnamefont {{Kuroda}}}, \bibinfo {author} {\bibfnamefont
  {T.}~\bibnamefont {{Kondo}}}, \bibinfo {author} {\bibfnamefont
  {K.}~\bibnamefont {{Okazaki}}}, \bibinfo {author} {\bibfnamefont
  {K.}~\bibnamefont {{Kindo}}}, \bibinfo {author} {\bibfnamefont
  {X.}~\bibnamefont {{Wang}}}, \bibinfo {author} {\bibfnamefont
  {C.}~\bibnamefont {{Jin}}}, \bibinfo {author} {\bibfnamefont
  {J.}~\bibnamefont {{Hu}}}, \bibinfo {author} {\bibfnamefont {R.}~\bibnamefont
  {{Thomale}}}, \bibinfo {author} {\bibfnamefont {K.}~\bibnamefont {{Sumida}}},
  \bibinfo {author} {\bibfnamefont {S.}~\bibnamefont {{Wu}}}, \bibinfo {author}
  {\bibfnamefont {K.}~\bibnamefont {{Miyamoto}}}, \bibinfo {author}
  {\bibfnamefont {T.}~\bibnamefont {{Okuda}}}, \bibinfo {author} {\bibfnamefont
  {H.}~\bibnamefont {{Ding}}}, \bibinfo {author} {\bibfnamefont {G.~D.}\
  \bibnamefont {{Gu}}}, \bibinfo {author} {\bibfnamefont {T.}~\bibnamefont
  {{Tamegai}}}, \bibinfo {author} {\bibfnamefont {T.}~\bibnamefont
  {{Kawakami}}}, \bibinfo {author} {\bibfnamefont {M.}~\bibnamefont {{Sato}}},\
  and\ \bibinfo {author} {\bibfnamefont {S.}~\bibnamefont {{Shin}}},\
  }\bibfield  {title} {\bibinfo {title} {{Multiple topological states in
  iron-based superconductors}},\ }\href
  {https://doi.org/10.1038/s41567-018-0280-z} {\bibfield  {journal} {\bibinfo
  {journal} {Nature Physics}\ }\textbf {\bibinfo {volume} {15}},\ \bibinfo
  {pages} {41} (\bibinfo {year} {2019})},\ \Eprint
  {https://arxiv.org/abs/1809.09977} {arXiv:1809.09977 [cond-mat.supr-con]}
  \BibitemShut {NoStop}%
\bibitem [{\citenamefont {Oudah}\ \emph {et~al.}(2016)\citenamefont {Oudah},
  \citenamefont {Ikeda}, \citenamefont {Hausmann}, \citenamefont {Yonezawa},
  \citenamefont {Fukumoto}, \citenamefont {Kobayashi}, \citenamefont {Sato},\
  and\ \citenamefont {Maeno}}]{oudah2016dirac}%
  \BibitemOpen
  \bibfield  {author} {\bibinfo {author} {\bibfnamefont {M.}~\bibnamefont
  {Oudah}}, \bibinfo {author} {\bibfnamefont {A.}~\bibnamefont {Ikeda}},
  \bibinfo {author} {\bibfnamefont {J.~N.}\ \bibnamefont {Hausmann}}, \bibinfo
  {author} {\bibfnamefont {S.}~\bibnamefont {Yonezawa}}, \bibinfo {author}
  {\bibfnamefont {T.}~\bibnamefont {Fukumoto}}, \bibinfo {author}
  {\bibfnamefont {S.}~\bibnamefont {Kobayashi}}, \bibinfo {author}
  {\bibfnamefont {M.}~\bibnamefont {Sato}},\ and\ \bibinfo {author}
  {\bibfnamefont {Y.}~\bibnamefont {Maeno}},\ }\bibfield  {title} {\bibinfo
  {title} {Superconductivity in the antiperovskite dirac-metal oxide sr3- x
  sno},\ }\href {https://doi.org/10.1038/ncomms13617} {\bibfield  {journal}
  {\bibinfo  {journal} {Nature communications}\ }\textbf {\bibinfo {volume}
  {7}},\ \bibinfo {pages} {13617} (\bibinfo {year} {2016})}\BibitemShut
  {NoStop}%
\bibitem [{\citenamefont {Schoop}\ \emph {et~al.}(2015)\citenamefont {Schoop},
  \citenamefont {Xie}, \citenamefont {Chen}, \citenamefont {Gibson},
  \citenamefont {Lapidus}, \citenamefont {Kimchi}, \citenamefont
  {Hirschberger}, \citenamefont {Haldolaarachchige}, \citenamefont {Ali},
  \citenamefont {Belvin}, \citenamefont {Liang}, \citenamefont {Neaton},
  \citenamefont {Ong}, \citenamefont {Vishwanath},\ and\ \citenamefont
  {Cava}}]{Schoop15}%
  \BibitemOpen
  \bibfield  {author} {\bibinfo {author} {\bibfnamefont {L.~M.}\ \bibnamefont
  {Schoop}}, \bibinfo {author} {\bibfnamefont {L.~S.}\ \bibnamefont {Xie}},
  \bibinfo {author} {\bibfnamefont {R.}~\bibnamefont {Chen}}, \bibinfo {author}
  {\bibfnamefont {Q.~D.}\ \bibnamefont {Gibson}}, \bibinfo {author}
  {\bibfnamefont {S.~H.}\ \bibnamefont {Lapidus}}, \bibinfo {author}
  {\bibfnamefont {I.}~\bibnamefont {Kimchi}}, \bibinfo {author} {\bibfnamefont
  {M.}~\bibnamefont {Hirschberger}}, \bibinfo {author} {\bibfnamefont
  {N.}~\bibnamefont {Haldolaarachchige}}, \bibinfo {author} {\bibfnamefont
  {M.~N.}\ \bibnamefont {Ali}}, \bibinfo {author} {\bibfnamefont {C.~A.}\
  \bibnamefont {Belvin}}, \bibinfo {author} {\bibfnamefont {T.}~\bibnamefont
  {Liang}}, \bibinfo {author} {\bibfnamefont {J.~B.}\ \bibnamefont {Neaton}},
  \bibinfo {author} {\bibfnamefont {N.~P.}\ \bibnamefont {Ong}}, \bibinfo
  {author} {\bibfnamefont {A.}~\bibnamefont {Vishwanath}},\ and\ \bibinfo
  {author} {\bibfnamefont {R.~J.}\ \bibnamefont {Cava}},\ }\bibfield  {title}
  {\bibinfo {title} {Dirac metal to topological metal transition at a
  structural phase change in ${\mathrm{au}}_{2}\mathrm{Pb}$ and prediction of
  ${\mathbb{z}}_{2}$ topology for the superconductor},\ }\href
  {https://doi.org/10.1103/PhysRevB.91.214517} {\bibfield  {journal} {\bibinfo
  {journal} {Phys. Rev. B}\ }\textbf {\bibinfo {volume} {91}},\ \bibinfo
  {pages} {214517} (\bibinfo {year} {2015})}\BibitemShut {NoStop}%
\bibitem [{\citenamefont {Kim}\ \emph {et~al.}(2018)\citenamefont {Kim},
  \citenamefont {Wang}, \citenamefont {Nakajima}, \citenamefont {Hu},
  \citenamefont {Ziemak}, \citenamefont {Syers}, \citenamefont {Wang},
  \citenamefont {Hodovanets}, \citenamefont {Denlinger}, \citenamefont {Brydon}
  \emph {et~al.}}]{kim2018beyond}%
  \BibitemOpen
  \bibfield  {author} {\bibinfo {author} {\bibfnamefont {H.}~\bibnamefont
  {Kim}}, \bibinfo {author} {\bibfnamefont {K.}~\bibnamefont {Wang}}, \bibinfo
  {author} {\bibfnamefont {Y.}~\bibnamefont {Nakajima}}, \bibinfo {author}
  {\bibfnamefont {R.}~\bibnamefont {Hu}}, \bibinfo {author} {\bibfnamefont
  {S.}~\bibnamefont {Ziemak}}, \bibinfo {author} {\bibfnamefont
  {P.}~\bibnamefont {Syers}}, \bibinfo {author} {\bibfnamefont
  {L.}~\bibnamefont {Wang}}, \bibinfo {author} {\bibfnamefont {H.}~\bibnamefont
  {Hodovanets}}, \bibinfo {author} {\bibfnamefont {J.~D.}\ \bibnamefont
  {Denlinger}}, \bibinfo {author} {\bibfnamefont {P.~M.}\ \bibnamefont
  {Brydon}}, \emph {et~al.},\ }\bibfield  {title} {\bibinfo {title} {Beyond
  triplet: Unconventional superconductivity in a spin-3/2 topological
  semimetal},\ }\href {https://www.science.org/doi/10.1126/sciadv.aao4513}
  {\bibfield  {journal} {\bibinfo  {journal} {Science advances}\ }\textbf
  {\bibinfo {volume} {4}},\ \bibinfo {pages} {eaao4513} (\bibinfo {year}
  {2018})}\BibitemShut {NoStop}%
\bibitem [{\citenamefont {Ishihara}\ \emph {et~al.}(2021)\citenamefont
  {Ishihara}, \citenamefont {Takenaka}, \citenamefont {Miao}, \citenamefont
  {Mizukami}, \citenamefont {Hashimoto}, \citenamefont {Yamashita},
  \citenamefont {Konczykowski}, \citenamefont {Masuki}, \citenamefont
  {Hirayama}, \citenamefont {Nomoto}, \citenamefont {Arita}, \citenamefont
  {Pavlosiuk}, \citenamefont {Wi\ifmmode~\acute{s}\else \'{s}\fi{}niewski},
  \citenamefont {Kaczorowski},\ and\ \citenamefont
  {Shibauchi}}]{ishihara2021heusler}%
  \BibitemOpen
  \bibfield  {author} {\bibinfo {author} {\bibfnamefont {K.}~\bibnamefont
  {Ishihara}}, \bibinfo {author} {\bibfnamefont {T.}~\bibnamefont {Takenaka}},
  \bibinfo {author} {\bibfnamefont {Y.}~\bibnamefont {Miao}}, \bibinfo {author}
  {\bibfnamefont {Y.}~\bibnamefont {Mizukami}}, \bibinfo {author}
  {\bibfnamefont {K.}~\bibnamefont {Hashimoto}}, \bibinfo {author}
  {\bibfnamefont {M.}~\bibnamefont {Yamashita}}, \bibinfo {author}
  {\bibfnamefont {M.}~\bibnamefont {Konczykowski}}, \bibinfo {author}
  {\bibfnamefont {R.}~\bibnamefont {Masuki}}, \bibinfo {author} {\bibfnamefont
  {M.}~\bibnamefont {Hirayama}}, \bibinfo {author} {\bibfnamefont
  {T.}~\bibnamefont {Nomoto}}, \bibinfo {author} {\bibfnamefont
  {R.}~\bibnamefont {Arita}}, \bibinfo {author} {\bibfnamefont
  {O.}~\bibnamefont {Pavlosiuk}}, \bibinfo {author} {\bibfnamefont
  {P.}~\bibnamefont {Wi\ifmmode~\acute{s}\else \'{s}\fi{}niewski}}, \bibinfo
  {author} {\bibfnamefont {D.}~\bibnamefont {Kaczorowski}},\ and\ \bibinfo
  {author} {\bibfnamefont {T.}~\bibnamefont {Shibauchi}},\ }\bibfield  {title}
  {\bibinfo {title} {Tuning the parity mixing of singlet-septet pairing in a
  half-heusler superconductor},\ }\href
  {https://doi.org/10.1103/PhysRevX.11.041048} {\bibfield  {journal} {\bibinfo
  {journal} {Phys. Rev. X}\ }\textbf {\bibinfo {volume} {11}},\ \bibinfo
  {pages} {041048} (\bibinfo {year} {2021})}\BibitemShut {NoStop}%
\bibitem [{\citenamefont {Zhang}\ \emph {et~al.}(2019)\citenamefont {Zhang},
  \citenamefont {Yin}, \citenamefont {Dai}, \citenamefont {Zheng},
  \citenamefont {Chang}, \citenamefont {Belopolski}, \citenamefont {Wang},
  \citenamefont {Lin}, \citenamefont {Wang}, \citenamefont {Jin},\ and\
  \citenamefont {Hasan}}]{Zhang19b}%
  \BibitemOpen
  \bibfield  {author} {\bibinfo {author} {\bibfnamefont {S.~S.}\ \bibnamefont
  {Zhang}}, \bibinfo {author} {\bibfnamefont {J.-X.}\ \bibnamefont {Yin}},
  \bibinfo {author} {\bibfnamefont {G.}~\bibnamefont {Dai}}, \bibinfo {author}
  {\bibfnamefont {H.}~\bibnamefont {Zheng}}, \bibinfo {author} {\bibfnamefont
  {G.}~\bibnamefont {Chang}}, \bibinfo {author} {\bibfnamefont
  {I.}~\bibnamefont {Belopolski}}, \bibinfo {author} {\bibfnamefont
  {X.}~\bibnamefont {Wang}}, \bibinfo {author} {\bibfnamefont {H.}~\bibnamefont
  {Lin}}, \bibinfo {author} {\bibfnamefont {Z.}~\bibnamefont {Wang}}, \bibinfo
  {author} {\bibfnamefont {C.}~\bibnamefont {Jin}},\ and\ \bibinfo {author}
  {\bibfnamefont {M.~Z.}\ \bibnamefont {Hasan}},\ }\bibfield  {title} {\bibinfo
  {title} {Vector field controlled vortex lattice symmetry in lifeas using
  scanning tunneling microscopy},\ }\href
  {https://doi.org/10.1103/PhysRevB.99.161103} {\bibfield  {journal} {\bibinfo
  {journal} {Phys. Rev. B}\ }\textbf {\bibinfo {volume} {99}},\ \bibinfo
  {pages} {161103} (\bibinfo {year} {2019})}\BibitemShut {NoStop}%
\bibitem [{\citenamefont {Hoffmann}\ \emph {et~al.}(2022)\citenamefont
  {Hoffmann}, \citenamefont {Schlegel}, \citenamefont {Salazar}, \citenamefont
  {Sykora}, \citenamefont {Nag}, \citenamefont {Khanenko}, \citenamefont
  {Beck}, \citenamefont {Aswartham}, \citenamefont {Wurmehl}, \citenamefont
  {B\"uchner}, \citenamefont {Fasano},\ and\ \citenamefont
  {Hess}}]{Hoffmann22}%
  \BibitemOpen
  \bibfield  {author} {\bibinfo {author} {\bibfnamefont {S.}~\bibnamefont
  {Hoffmann}}, \bibinfo {author} {\bibfnamefont {R.}~\bibnamefont {Schlegel}},
  \bibinfo {author} {\bibfnamefont {C.}~\bibnamefont {Salazar}}, \bibinfo
  {author} {\bibfnamefont {S.}~\bibnamefont {Sykora}}, \bibinfo {author}
  {\bibfnamefont {P.~K.}\ \bibnamefont {Nag}}, \bibinfo {author} {\bibfnamefont
  {P.}~\bibnamefont {Khanenko}}, \bibinfo {author} {\bibfnamefont
  {R.}~\bibnamefont {Beck}}, \bibinfo {author} {\bibfnamefont {S.}~\bibnamefont
  {Aswartham}}, \bibinfo {author} {\bibfnamefont {S.}~\bibnamefont {Wurmehl}},
  \bibinfo {author} {\bibfnamefont {B.}~\bibnamefont {B\"uchner}}, \bibinfo
  {author} {\bibfnamefont {Y.}~\bibnamefont {Fasano}},\ and\ \bibinfo {author}
  {\bibfnamefont {C.}~\bibnamefont {Hess}},\ }\bibfield  {title} {\bibinfo
  {title} {Absence of hexagonal-to-square lattice transition in lifeas vortex
  matter},\ }\href {https://doi.org/10.1103/PhysRevB.106.134507} {\bibfield
  {journal} {\bibinfo  {journal} {Phys. Rev. B}\ }\textbf {\bibinfo {volume}
  {106}},\ \bibinfo {pages} {134507} (\bibinfo {year} {2022})}\BibitemShut
  {NoStop}%
\bibitem [{\citenamefont {Li}\ \emph {et~al.}(2022)\citenamefont {Li},
  \citenamefont {Li}, \citenamefont {Cao}, \citenamefont {Zhou}, \citenamefont
  {Wang}, \citenamefont {Jin}, \citenamefont {Chiu}, \citenamefont {Pennycook},
  \citenamefont {Wang},\ and\ \citenamefont {Gao}}]{li2022ordered}%
  \BibitemOpen
  \bibfield  {author} {\bibinfo {author} {\bibfnamefont {M.}~\bibnamefont
  {Li}}, \bibinfo {author} {\bibfnamefont {G.}~\bibnamefont {Li}}, \bibinfo
  {author} {\bibfnamefont {L.}~\bibnamefont {Cao}}, \bibinfo {author}
  {\bibfnamefont {X.}~\bibnamefont {Zhou}}, \bibinfo {author} {\bibfnamefont
  {X.}~\bibnamefont {Wang}}, \bibinfo {author} {\bibfnamefont {C.}~\bibnamefont
  {Jin}}, \bibinfo {author} {\bibfnamefont {C.-K.}\ \bibnamefont {Chiu}},
  \bibinfo {author} {\bibfnamefont {S.~J.}\ \bibnamefont {Pennycook}}, \bibinfo
  {author} {\bibfnamefont {Z.}~\bibnamefont {Wang}},\ and\ \bibinfo {author}
  {\bibfnamefont {H.-J.}\ \bibnamefont {Gao}},\ }\bibfield  {title} {\bibinfo
  {title} {Ordered and tunable majorana-zero-mode lattice in naturally strained
  lifeas},\ }\href {https://doi.org/10.1038/s41586-022-04744-8} {\bibfield
  {journal} {\bibinfo  {journal} {Nature}\ }\textbf {\bibinfo {volume} {606}},\
  \bibinfo {pages} {890} (\bibinfo {year} {2022})}\BibitemShut {NoStop}%
\bibitem [{\citenamefont {Lapp}\ \emph {et~al.}(2020)\citenamefont {Lapp},
  \citenamefont {B\"orner},\ and\ \citenamefont {Timm}}]{lapp2020BdG}%
  \BibitemOpen
  \bibfield  {author} {\bibinfo {author} {\bibfnamefont {C.~J.}\ \bibnamefont
  {Lapp}}, \bibinfo {author} {\bibfnamefont {G.}~\bibnamefont {B\"orner}},\
  and\ \bibinfo {author} {\bibfnamefont {C.}~\bibnamefont {Timm}},\ }\bibfield
  {title} {\bibinfo {title} {Experimental consequences of bogoliubov fermi
  surfaces},\ }\href {https://doi.org/10.1103/PhysRevB.101.024505} {\bibfield
  {journal} {\bibinfo  {journal} {Phys. Rev. B}\ }\textbf {\bibinfo {volume}
  {101}},\ \bibinfo {pages} {024505} (\bibinfo {year} {2020})}\BibitemShut
  {NoStop}%
\bibitem [{\citenamefont {Fu}\ and\ \citenamefont {Kane}(2008)}]{Fu08}%
  \BibitemOpen
  \bibfield  {author} {\bibinfo {author} {\bibfnamefont {L.}~\bibnamefont
  {Fu}}\ and\ \bibinfo {author} {\bibfnamefont {C.~L.}\ \bibnamefont {Kane}},\
  }\bibfield  {title} {\bibinfo {title} {Superconducting proximity effect and
  majorana fermions at the surface of a topological insulator},\ }\href
  {https://doi.org/10.1103/PhysRevLett.100.096407} {\bibfield  {journal}
  {\bibinfo  {journal} {Phys. Rev. Lett.}\ }\textbf {\bibinfo {volume} {100}},\
  \bibinfo {pages} {096407} (\bibinfo {year} {2008})}\BibitemShut {NoStop}%
\end{thebibliography}%

\newpage 
\cleardoublepage

\onecolumngrid
\begin{center}
\textbf{\large Supplemental Material for ``Topologically protected emergent Fermi surface in an Abrikosov vortex lattice"}\\[5pt]
\vspace{0.1cm}
\end{center}

\begin{center}
 {\small Songyang Pu$^{1,2}$, Jay D. Sau$^{3}$, Rui-Xing Zhang$^{1,2,4}$}  
\end{center}

\begin{center}
{\sl \footnotesize
$^{1}$Department of Physics and Astronomy, The University of Tennessee, Knoxville, TN 37996, USA

$^{2}$Institute of Advanced Materials and Manufacturing, The University of Tennessee, Knoxville, TN 37920, USA

$^{3}$Condensed Matter Theory Center and Joint Quantum Institute, Department of Physics, University of Maryland,
College Park, Maryland 20742, USA

$^{4}$Department of Materials Science and Engineering, The University of Tennessee, Knoxville, TN 37996, USA
}
\end{center}

\begin{center}
\begin{quote}
{\small In this supplemental material, we include an analytical solution of the nodal vortex state in a superconducting Dirac semimetal, a discussion of the Peierls phase in both effective model and microscopic model, detailed discussions on both the effective model and the microscopic model regarding the symmetry and topological properties, a discussion on the effect of vortex lattice constant on the CdGM band dispersions and topological phase diagrams, and additional numerical studies of the CdGM Fermi surface in the momentum space.
}\\[20pt]
\end{quote}
\end{center}

\tableofcontents

\setcounter{equation}{0}
\setcounter{figure}{0}
\setcounter{table}{0}
\setcounter{page}{1}
\setcounter{section}{0}
\makeatletter
\renewcommand{\theequation}{S\arabic{equation}}
\renewcommand{\thefigure}{S\arabic{figure}}
\renewcommand{\thesection}{S\Roman{section}}
\renewcommand{\thepage}{\arabic{page}}
\renewcommand{\thetable}{S\arabic{table}}

\vspace{0cm}
\section{Appendix A: Nodal Vortex in a Superconducting Dirac Semimetal}

In this appendix, we review the origin of the nodal vortex in 3D superconducting Dirac semimetals (DSMs). We will focus on a rotation-symmetry-protected 3D DSM with $s$-wave spin-singlet pairing in the BdG formalism. By including a vortex phase winding, we will analytically solve for the 1D dispersion relation of vortex-line bound states. A similar derivation can be found in Ref.~\cite{Konig19}. The basis functions for the normal-state DSM is
\begin{equation}
|\Psi\rangle = (|\frac{1}{2}\rangle, |-\frac{1}{2}\rangle, |\frac{3}{2}\rangle, |-\frac{3}{2}\rangle)^T,
\end{equation}
where $|J_z\rangle$ labels the $\hat{z}$-component angular momentum. Under this basis choice, the DSM Hamiltonian is 
\begin{equation}
h_0 = \begin{pmatrix}
M(k) & 0 & v k_+ & 0 \\
0 & M(k) & 0 & vk_- \\
v k_- & 0 & -M(k) & 0 \\
0 & -vk_+ & 0 & -M(k)
\end{pmatrix},
\end{equation}
where $M(k)=M_0-M_z k_z^2 - M_\parallel (k_x^2 + k_y^2)$. We choose the Nambu basis
\begin{eqnarray}
|\Psi\rangle & = & |\Psi_+\rangle \oplus |\Psi_-\rangle \nonumber \\
& = & (|\frac{1}{2},e\rangle, |\frac{3}{2},e\rangle, |\frac{1}{2},h\rangle, |\frac{3}{2},h\rangle)^T \oplus  (|-\frac{1}{2},e\rangle, |-\frac{3}{2},e\rangle, -|-\frac{1}{2},h\rangle, -|-\frac{3}{2},h\rangle)^T.
\end{eqnarray}
Under this basis, the BdG Hamiltonian admits a block-diagonal form $H=H_+ \oplus H_-$, where
\begin{eqnarray}
H_+ &=& \begin{pmatrix}
M(k)-\mu & vk_+ & -\Delta(r) e^{i\theta} & 0 \\
v k_- & -M(k)-\mu & 0 & -\Delta e^{i\theta} \\
-\Delta e^{-i\theta} & 0 & -M(k)+\mu & -vk_+ \\
0 & -\Delta e^{-i\theta} & -v k_- & M(k)+\mu \\
\end{pmatrix} \nonumber \\
H_- &=& \begin{pmatrix}
M(k)-\mu & -vk_- & -\Delta(r) e^{i\theta} & 0 \\
-v k_+ & -M(k)-\mu & 0 & -\Delta e^{i\theta} \\
-\Delta e^{-i\theta} & 0 & -M(k)+\mu & vk_- \\
0 & -\Delta e^{-i\theta} & v k_+ & M(k)+\mu \\
\end{pmatrix}.
\end{eqnarray}
It is easy to check that the particle-hole symmetry (PHS) now becomes
\begin{equation}
\Xi = \tau_y \otimes \sigma_y \otimes s_0 {\cal K} = \begin{pmatrix}
& & & -s_0 \\
& & s_0 & \\
& s_0 & & \\
-s_0 & & & \\
\end{pmatrix} {\cal K},
\end{equation}
which switches between $H_{\pm}$. Since
\begin{equation}
	H_{\pm} = M(k) \sigma_z \otimes s_z \pm v (k_x \sigma_z \otimes s_x \mp k_y \sigma_z \otimes s_y) - \Delta (\cos\theta \sigma_x - \sin \theta \sigma_y) - \mu \sigma_z,
\end{equation}
we note that both $H_{\pm}$ feature an emergent particle hole symmetry $\widetilde{\Xi} = \sigma_y \otimes s_y {\cal K}$ so that 
\begin{equation}
	\widetilde{\Xi} H^{(0)}_{\pm}(k) \widetilde{\Xi}^{-1}  = -H^{(0)}_{\pm} (-k)^T
\end{equation}
with $H^{(0)}_{\pm} (k) = H_{\pm} (k) - M(k) \sigma_z \otimes s_z$. Notably, $H_\pm^{(0)}$ is exactly the known Fu-Kane Hamiltonian for vortex Majorana bound states on the surface of a superconducting topological insulator~\cite{Fu08}.\\

Since $H_+$ and $H_-$ are related to each other by PHS, we only need to solve the zero-energy vortex wavefunction for $H_+$. In the polar coordinate $r=\sqrt{x^2+y^2}$ and $\theta = \tan^{-1} (y/x)$, we have
\begin{equation}
	k_\pm = -i e^{\pm i \theta} (\partial_r \pm  \frac{i}{r} \partial_{\theta}).
\end{equation} 
For our purpose, we take $M(k)$ term as a small perturbation, and the zeroth-order Hamiltonian is 
\begin{equation}
	H_+^{(0)} = \begin{pmatrix}
	-\mu & -iv e^{ i \theta} (\partial_r + \frac{i}{r} \partial_{\theta}) & -\Delta(r) e^{i\theta} & 0 \\
	-iv e^{-i \theta} (\partial_r - \frac{i}{r} \partial_{\theta}) & -\mu & 0 & -\Delta e^{i\theta} \\
	-\Delta e^{-i\theta} & 0 & \mu & iv e^{i \theta} (\partial_r + \frac{i}{r} \partial_{\theta}) \\
	0 & -\Delta e^{-i\theta} & iv e^{-i \theta} (\partial_r - \frac{i}{r} \partial_{\theta}) & \mu \\
	\end{pmatrix}.
\end{equation}
Consider a trial wavefunction 
\begin{equation}
	|\Psi_+^{(l_z)} \rangle = [A(r)e^{il_z\theta}, B(r)e^{i(l_z-1)\theta}, C(r)e^{i(l_z-1)\theta}, D(r)e^{i(l_z-2)\theta}]^T
\end{equation}
The zero-mode equation is
\begin{eqnarray}
	H_+^{(0)} |\Psi_+^{(l_z)} \rangle = 0,
\end{eqnarray}
which is equivalent to 
\begin{eqnarray}
	{\cal H}_{l_z}\begin{pmatrix}
	A(r) \\
	B(r) \\
	C(r) \\
	D(r) \\
	\end{pmatrix} =
	\begin{pmatrix}
	-\mu & -iv (\partial_r - \frac{l_z-1}{r}) & -\Delta(r) & 0 \\
	-iv (\partial_r + \frac{l_z}{r}) & -\mu & 0 & -\Delta \\
	-\Delta  & 0 & \mu & iv (\partial_r - \frac{l_z-2}{r}) \\
	0 & -\Delta  & iv (\partial_r + \frac{l_z-1}{r}) & \mu \\
	\end{pmatrix} 
	\begin{pmatrix}
	A(r) \\
	B(r) \\
	C(r) \\
	D(r) \\
	\end{pmatrix} = 0.
\end{eqnarray}
It is important to note that 
\begin{equation}
	{\cal H}_{l_z} = \mu \tau_z \otimes \sigma_0 -iv \partial_r \tau_z \otimes \sigma_x - \Delta \tau_x \otimes \sigma_0 - \frac{v(l_z-1)}{r} \tau_z \otimes \sigma_y + \frac{iv}{r} (i\tau_0\otimes \sigma_y - \tau_z \otimes \sigma_x),
\end{equation}
and admits a chiral symmetry ${\cal S} = \tau_y \otimes \sigma_x$ only when $l=1$. Therefore, a zero-mode solution is only possible when $l_z=1$.

Such a zero mode must be simultaneously an eigenstate of ${\cal S}$, and thus satisfies
\begin{equation}
	{\cal S} \begin{pmatrix}
	A(r) \\
	B(r) \\
	C(r) \\
	D(r) \\
	\end{pmatrix} = \lambda \begin{pmatrix}
	A(r) \\
	B(r) \\
	C(r) \\
	D(r) \\
	\end{pmatrix},
\end{equation}
with $\lambda=\pm$. As a result, we have
\begin{equation}
	C(r) = i\lambda B(r),\ \ D(r) = i\lambda A(r),
\end{equation}
and the zero-mode equation can be simplified to 
\begin{equation}
	\begin{pmatrix}
	-\mu & -v (\partial_r - \frac{l_z-1}{r}) -\lambda \Delta(r) \\
	-v (\partial_r + \frac{l_z}{r}) - \lambda\Delta & -\mu \\
	\end{pmatrix} 
	\begin{pmatrix}
	e^{i\frac{\pi}{4}}A(r) \\
	e^{-i\frac{\pi}{4}}B(r) \\
	\end{pmatrix} = 0
\end{equation}
Notice that the Bessel function of the first kind $J_m(r)$ satisfies
\begin{equation}
	(\partial_r + \frac{m+1}{r}) J_{m+1}(k_F r) = k_F J_m(k_F r),
\end{equation}
where we have defined the Fermi momentum $k_F = \frac{\mu}{v}$. Then we have 
\begin{eqnarray}
	A(r) &=& e^{-i\frac{\pi}{4}} J_1(k_F r) f(r), \nonumber \\
	B(r) &=& e^{i\frac{\pi}{4}} J_0(k_F r) f(r),
\end{eqnarray}
where 
\begin{equation}
	f(r) = e^{-\int_0^r \frac{\Delta(r')}{v} dr'}
\end{equation}
and we have chosen $\lambda=+1$ to guarantee the normalizability of the wavefunction. We now conclude that the zero-mode wavefunction for $H_+$ is 
\begin{equation}
	\Psi_{l_z=1}(r,\theta) = {\cal N} e^{-i\frac{\pi}{4}}
	\begin{pmatrix}
	J_1(k_F r) e^{i\theta} \\
	i J_0(k_F r) \\
	- J_0 (k_F r) \\
	i J_1(k_F r) e^{-i\theta} \\
	\end{pmatrix} f(r)
\end{equation}
For the superconducting order parameter, we take
\begin{equation}
	\Delta(r) = \Delta_0 \tanh \frac{r}{\xi_0},
\end{equation}
where $\xi_0$ is the superconducting coherence length. Then we have
\begin{equation}
	f(r) = (\cosh \frac{r}{\xi_0})^{-\frac{\xi_0}{\xi}},
\end{equation}
with $\xi = \frac{v}{\Delta_0}$. In the asymptotic limit, we have
\begin{equation}
	f(r) \approx e^{-\frac{r}{\xi}},
\end{equation}
which is independent of the SC coherence length $\xi_0$. \\

Since $H_-$ is related to $H_+$ by PHS, its zero-mode wavefunction is thus given by
\begin{equation}
	\Psi_{l_z=-1}(r,\theta) = {\cal N} \Xi \Psi_{l_z=1}(r,\theta) = e^{i\frac{\pi}{4}}
	\begin{pmatrix}
	-J_0(k_F r) \\
	-i J_1(k_F r) e^{i\theta} \\
	- J_1 (k_F r) e^{-i\theta} \\
	i J_0(k_F r) \\
	\end{pmatrix} f(r)
\end{equation}
The normalization factor ${\cal N}$ for both wavefunctions is
\begin{equation}
	{\cal N} = \sqrt{\frac{1}{4\pi\int_0^{\infty} [J_0^2(k_F r) + J_1^2(k_F r)] dr}}.
\end{equation}
It should be noted that the angular momentum operator for the vortex mode is
\begin{equation}
	l_z = -i \partial_{\theta} + L_z - \frac{1}{2} \tau_z.
\end{equation}
Here $L_z$ is the contribution of angular momentum from the atomic basis. The factor $-\frac{1}{2}\tau_z$ implies a shift of $-\frac{1}{2}$ ($\frac{1}{2}$) for the angular momentum for electron (hole) states, which originates from the $\pi$-flux of the superconducting vortex. It is then easy to see that $L_z$ is exactly the $l$-label of each wave function.

The effective Hamiltonian is obtained by projecting the $M(k)$ term onto the low-energy space spanned by $\Psi_{l=\pm1}$, which is 
\begin{equation}
	h_\text{eff}(k_z) = \alpha (M_0-M_z k_z^2)\sigma_z,
\end{equation}
where
\begin{eqnarray}
	\alpha = {\cal N}^2 m_0 \int_{0}^{\infty} [\chi_0^2(r) - \chi_1^2(r)] dr.
\end{eqnarray}
Here we have defined 
\begin{eqnarray}
	\chi_{j}(r) = J_{j}(k_F r) f(r)\quad j=0,1.
\end{eqnarray}
Clearly, when the normal state is a DSM with $M_0M_z>0$, the vortex state is also gapless with a pair of nodes at $k_z=\pm\sqrt{M_0/M_z}$.

\section{Appendix B. Peierls Phase}

\begin{figure}[t]
    \includegraphics[width=0.3\columnwidth]{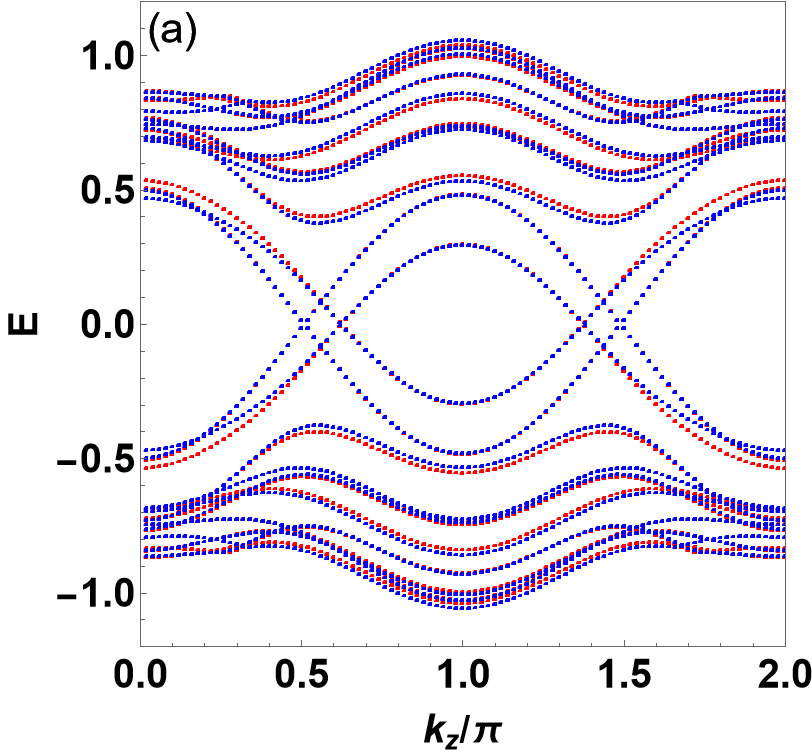}
    \includegraphics[width=0.3\columnwidth]{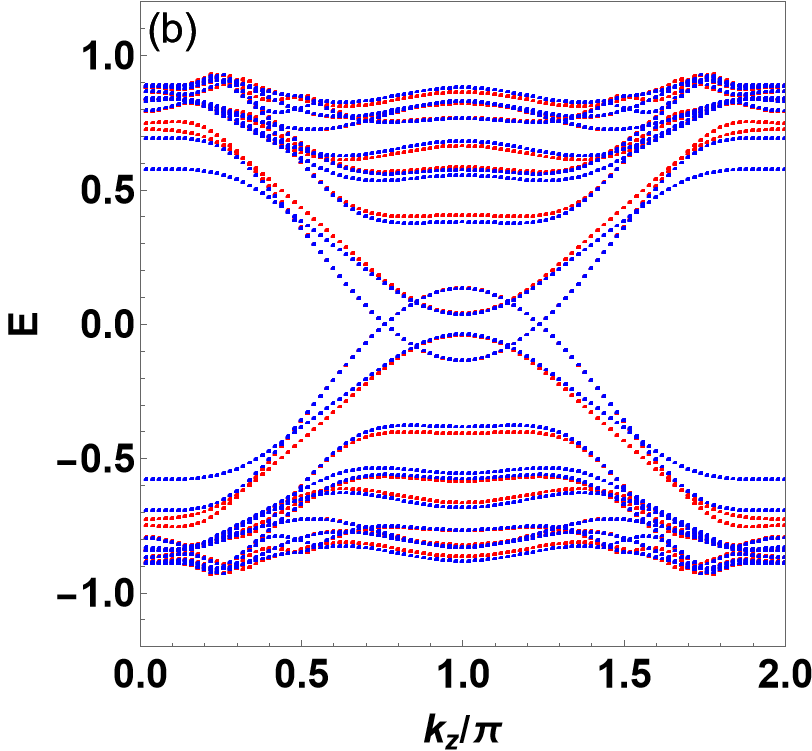}
    \caption{Band structures of the microscopic full model with different choices of cut-off $G_C=2L$ (blue) and $G_C=4L$ (red). We have chosen $M_0=1.8$ (a) and $M_0=1.45$(b). Other parameters are $L=6$, $M_1=v=2M_2=1$, $\mu=0.1$ and $\Delta_0=0.15$. }
    \label{cutoff}
\end{figure}

In this part, we prove that the Peierls phase between neighboring vortices $\zeta_{\mathbf{r}_1,\mathbf{r}_2}=\int_{\mathbf{r}_1}^{\mathbf{r}_2}\vec{\Omega}\cdot d\vec{l}=\pm \pi/2$ is exact. Since the vector potential is singular at the vortices, the above path integral has two parts $\zeta_{\mathbf{r}_1,\mathbf{r}_2} = \zeta^{(1)} + \zeta^{(2)}$. The first part $\zeta^{(1)}$ accounts for the phase accumulation from two infinitesimally small quarter-circles that go around $\mathbf{r}_1$ and $\mathbf{r}_2$ respectively. It is easy to show that $\zeta^{(1)}= \pm \pi/2$ since a full circle around a vortex has $\pi$ phase.

$\zeta^{(2)}$ takes into account the contributions from a straight line that connects $\mathbf{r}_1$ and $\mathbf{r}_2$. For a square vortex lattice, $\delta\vec{r}=\vec{r}-\vec{R}_{\alpha}$ path can go along $x$ axis, $y$ axis, $y=x$ axis, or $y=-x$ axis. Let us denote the integrand of $\zeta^{(2)}$ as $\vec{\omega}(\mathbf{k})={i {\bf k}\times \hat{z}\over 1+\lambda^2|{\bf k}|^2}e^{i{\bf k}\cdot \delta{\bf r}}$. (The index of $\alpha$ is omitted in this section.) If $\delta{\bf r}$ goes along $x$ axis, $\vec{\omega}(k_x,k_y)=-\vec{\omega}(k_x,-k_y)$ and the path integral becomes zero. Similarly, if $\delta\vec{r}=(0,y)$, $\vec{\omega}(k_x,k_y)=-\vec{\omega}(-k_x,k_y)$. If $\delta\vec{r}=(x,x)$, ${\omega}_{x,y}(k_x,k_y)=-{\omega}_{y,x}(k_y,k_x)$, and if $\delta\vec{r}=(x,-x)$, ${\omega}_{x,y}(k_x,k_y)={\omega}_{y,x}(-k_y,-k_x)$. Hence, $\zeta^{(2)}=0$ always holds and we conclude that the net Peierls phase $\zeta=\pm \frac{\pi}{2}$. This fact is crucial for constructing our effective model for the vortex lattice. 

On the other hand, our microscopic simulation for the full lattice model is independent of the above fact. Instead, we directly calculate the integral of $\vec{\Omega}$ through the following equivalent form of Eq. 2 in the main text
\begin{equation}
    \vec{\Omega}_\alpha({\bf r})={2\pi\over L^2}\lambda^2\sum_{\bf G}{i {\bf G}\times \hat{z}\over 1+\lambda^2|{\bf G}|^2}e^{i{\bf G}\cdot ({\bf r}-{\bf R}_{\alpha})}
    \label{eq:Omega_s}
\end{equation}
where $L$ is the vortex lattice constant. In Eq.~\ref{eq:Omega_s}, the integration over $\vec{k}$ is replaced by the summation over reciprocal vortex-lattice vector $\vec{G}={2\pi\over L}(m,n)$ as the summation over lattice sites $i$ is implicitly taken. In practice, we put a cutoff $-G_C\leq m,n\leq G_C$, $G_C=2L$ on the summation over $\vec{G}$. We found that there only exist some tiny quantitative differences for the vortex-lattice band structures calculated with $G_C=2L$ and $G_C=4L$, as shown in Fig.~\ref{cutoff}. This ensures that our calculations are sufficiently convergent to identify and characterize the CdGM Fermi surfaces.

\section{Appendix C: Effective Model of a Vortex Lattice}

As discussed in the main text, the effective Hamiltonian for the nodal vortex lattice is 
 \begin{equation}
    {\cal H}(\vec{k})= g({\bf k}_\parallel)\tilde{\tau}_0\otimes \sigma_0 +m(k_z)\tilde{\tau}_z \otimes \sigma_0,
\end{equation}
where $m(k_z)=(m_0-m_zk_z^2)$ and $g({\bf k}_\parallel)=2t_1\sin {k_x+k_y\over 2}\sigma_x-2t_1\cos {k_x-k_y\over 2}\sigma_y+2t_2(\sin k_y-\sin k_x)\sigma_z$. In this part, we will discuss two important symmetries of ${\cal H}$: the inversion symmetry ${\cal P}$ and the PHS $\Xi$. 

Due to the non-trivial gauge structure, the inversion center in the $k$ space is shifted from $(0,0,0)$ to $(\pi/2,-\pi/2,0)$. Specifically, the inversion symmetry ${\cal P}$ satisfies
\begin{eqnarray}
    \mathcal{P}=\tilde{\tau}_z\otimes e^{-i{\pi\over 2}\sigma_z},\ \ \mathcal{P}H(\vec{k})\mathcal{P}^{-1}=H(\pi-k_x,-\pi-k_y,-k_z).
\end{eqnarray}
Similarly, the PHS also features a shift of the $k$-flipping center, with 
\begin{eqnarray}
   \Xi=\tilde{\tau}_x\otimes \sigma_x {\cal K},\ \ \Xi H(\vec{k})\Xi^{-1}=-H^T(\pi-k_x,-\pi-k_y,-k_z).
\end{eqnarray} 
Notably, the PHS necessarily involves an interchange of the $A$ and $B$ vortex sites (i.e., $\sigma_x$) following the specific gauge choice in Ref.~\cite{Franz00}. 

Clearly, the Hamiltonian is invariant under the combined symmetry $W=\Xi I$, $WH(\vec{k})W^{-1}=-H^T(\vec{k})$.
The symmetry operation can be diagonalized by a unitary congruence $W=Q\Lambda Q^T$ where $\Lambda$ is a diagonal matrix. Ref.~\cite{Agterberg17} proposed a $\mathbb{Z}_2$ topological invariant for Fermi surfaces of systems that preserve such symmetry. It is defined for the anti-symmetrized matrix $\tilde{H}(\vec{k})=\Omega H(\vec{k}) \Omega^\dagger$ where $\Omega=\sqrt{\Lambda}^\dagger Q^\dagger$. The $\mathbb{Z}_2$ invariant $\eta$ is defined as $(-1)^\eta={\rm sgn}\left[{\rm Pf}[\tilde{H}(\vec{k}_1)]{\rm Pf}[\tilde{H}(\vec{k}_2)]\right]$, where $\vec{k}_{1,2}$ are on different sides of the Fermi surface. (If they are on the same side, $\eta=0$ is guaranteed.) When $\eta=1$, the values of Pfaffian ${\rm Pf}[\tilde{H}(\vec{k})]$ change signs across the Fermi surface, necessarily indicating a gap closure. We confirmed numerically that all of our CdGM Fermi surfaces have $\eta=1$, which is consistent with the fact that the Chern number is altered by two when crossing CdGM Fermi surfaces. Thereby, the CdGM Fermi surface is topological and protected by the $\mathbb{Z}_2$ invariant $\eta$ in centrosymmetric BdG systems.

\section{Appendix D: Microscopic Simulation of a Vortex Lattice}

\subsection{D1. Model}

In this appendix, we describe in details our microscopic simulations of the vortex lattice spectrum of a superconducting DSM. 
Our starting point is a BdG Hamiltonian
\begin{equation}
    H({\bf k}) = \begin{pmatrix}
        h_0({\bf k}) - \mu & \Delta \\
        \Delta^* & \mu -h_0({\bf k}) \\
    \end{pmatrix}.
\end{equation}
where $h_0({\bf k})=M({\bf k})s_0\otimes \kappa_z + v (\sin k_x s_z\otimes \kappa_x - \sin k_ys_0\otimes \kappa_y)$, $M({\bf k})=M_0 - M_1(\cos k_x + \cos k_y) - M_2\cos k_z$ and $\kappa_i$ are the Pauli matrices for orbital degrees of freedom. Following the strategy in Ref.~\cite{Franz00}, we perform a unitary transformation $H\rightarrow UHU^{-1}$, where
\begin{eqnarray}
    U=\begin{pmatrix}
    e^{-i\theta_A(\vec{r})}&0\\
    0&e^{i\theta_B(\vec{r})}
\end{pmatrix}
\end{eqnarray}
redistributes all vortex phases from SC gap parameters to the electrons and holes. Here, $\theta_\alpha =\sum_i {\rm arg}({\bf r}-{\bf R}_{i\alpha})$ arises from the vortex $\alpha$ at ${\bf R}_i$.
The lattice constant of the vortex lattice is denoted as $L$, in units of atomic lattice site. Then the full microscopic Hamiltonian is given by
\begin{align}
\mathcal{H}_m&=\sum_{r_x,r_y,k_z}\left(M_0\tau_3\otimes s_0\otimes \kappa_3-\mu \tau_3\otimes s_0\otimes \kappa_0+\Delta \tau_1\otimes s_0\otimes \kappa_0\right)c_{\vec{r},k_z}^{\dagger}c_{\vec{r},k_z}\\ 
&+\left(-{1\over 2}M_1T_{\vec{\delta}_x}(\vec{r})\otimes s_0\otimes \kappa_3+{1\over 2i}vT_{\vec{\delta}_x}(\vec{r})\otimes s_3\otimes \kappa_1\right)c_{\vec{r}+\vec{\delta}_x,k_z}^{\dagger}c_{\vec{r},k_z}\\ 
&+\left(-{1\over 2}M_1T_{-\vec{\delta}_x}(\vec{r})\otimes s_0\otimes \kappa_3-{1\over 2i}vT_{-\vec{\delta}_x}(\vec{r})\otimes s_3\otimes \kappa_1\right)c_{\vec{r}-\vec{\delta}_x,k_z}^{\dagger}c_{\vec{r},k_z}\\
&+\left(-{1\over 2}M_1T_{\vec{\delta}_y}(\vec{r})\otimes s_0\otimes \kappa_3-{1\over 2i}vT_{\vec{\delta}_y}(\vec{r})\otimes s_3\otimes \kappa_1\right)c_{\vec{r}+\vec{\delta}_y,k_z}^{\dagger}c_{\vec{r},k_z}\\ 
&+\left(-{1\over 2}M_1T_{-\vec{\delta}_y}(\vec{r})\otimes s_0\otimes \kappa_3+{1\over 2i}vT_{-\vec{\delta}_x}(\vec{r})\otimes s_3\otimes \kappa_1\right)c_{\vec{r}-\vec{\delta}_y,k_z}^{\dagger}c_{\vec{r},k_z}
\end{align}
under the basis of
\begin{align}
|\Psi\rangle=&\left(|s\uparrow\rangle_e,|p\uparrow\rangle_e,|s\downarrow\rangle_e,|p\downarrow\rangle_e\right)^T\oplus \\ \nonumber
&\left(|s\downarrow\rangle_h,|p\downarrow\rangle_h,-|s\uparrow\rangle_h,-|p\uparrow\rangle_h\right)^T.
\end{align}
Here we define $\vec{r}=(r_x,r_y)$ as the atomic lattice sites, $\vec{\delta}_{x,y}=(1,0),(0,1)$ the nearest neighbor vector, and the two vortices are located at $(L/4,L/4)$ and $(3L/4,3L/4)$. Note that we have assumed the periodic boundary condition for the vortex unit cell. The Peirels phase factor of ${\bf \Omega}$ field is given by
\begin{equation}
T_{\vec{\delta}}(\vec{r})=\begin{pmatrix}
    e^{-i\int_{{\bm r}}^{{\bm r}+{\bm \delta}}\vec{\Omega}_A({\bm l})d{\bm l}} &0\\
    0&-e^{i\int_{{\bm r}}^{{\bm r}+{\bm \delta}}\vec{\Omega}_B({\bm l})d{\bm l}}
\end{pmatrix}.
\end{equation}

\subsection{D2. Symmetries}

We now discuss both PHS and inversion symmetry of ${\cal H}_m$. We find that the inversion symmetry is represented by
\begin{equation}
    \mathcal{P}_{i,j;i',j'}=\delta_{i+i',L-1}\delta_{j+j',L-1}\tau_0\otimes s_0\otimes \kappa_z 
\end{equation}
where $i,j,i',j'=0,1\cdots L-1$ are lattice site indices. The two delta-function factors in the above expression are because the inversion center coincides with the center of the vortex unit cell. The representation for the particle-hole symmetry can be written as 
\begin{equation}
    \Xi_{i,j;i',j'}=\delta_{i+L/2,i'}\delta_{j+L/2,j'}\tau_y\otimes s_y\otimes \kappa_0.
\end{equation}
We note that $\Xi$ contains a spatial translation by $(L/2,L/2)$, which arises from the gauge we choose. As we distribute all phases from vortex A to electrons and all phases from vortex B to holes, it is necessary to exchange A and B after particle-hole transformation, which is done by this translation. The $\mathbb{Z}_2$ invariant $\eta$ can be defined and numerically evaluated following the exactly same strategy for the effective model.

\begin{figure}[t]
    \includegraphics[width=0.3\columnwidth]{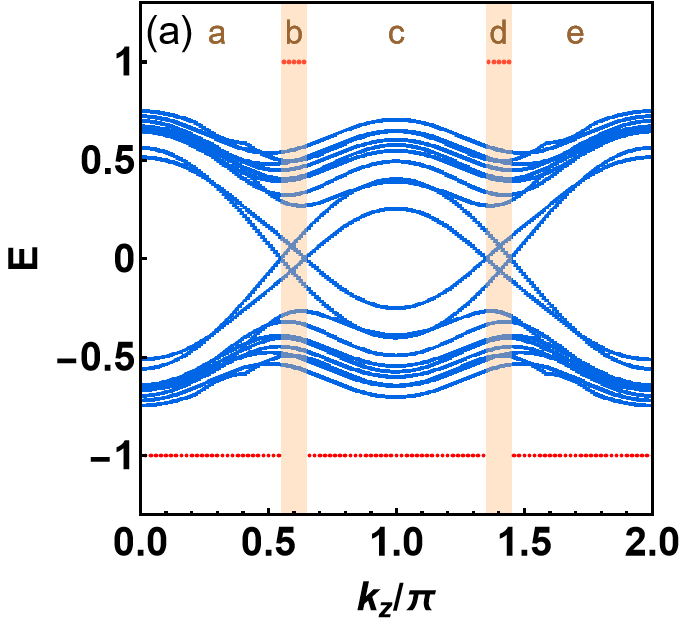}
    \includegraphics[width=0.3\columnwidth]{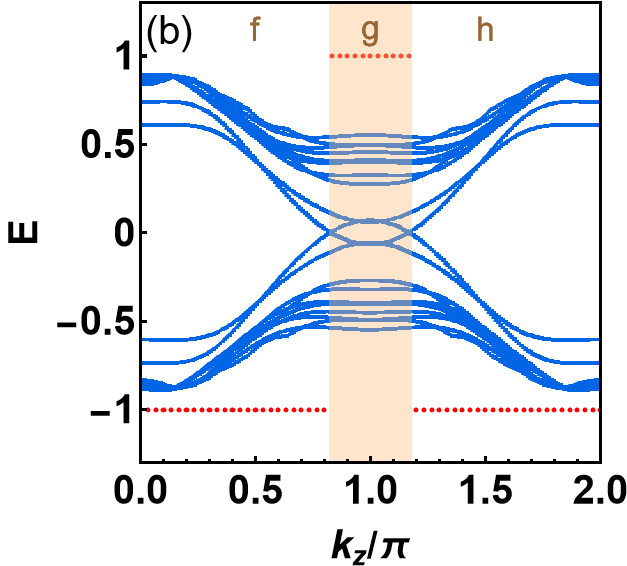}
    \caption{Energy dispersion and the value of ${\rm sgn}({\rm Pf}[\tilde{H}(\vec{k})])$ along the $k_z$ axis for $M_0=1.8$ in (a) with $\nu=2$ and $M_0=1.45$ in (b) with $\nu=1$.  We choose $L=10$, $M_1=v=2M_2=1$, $\mu=0.1$ and $\Delta_0=0.15$. The brown letters (e.g., a, b, c,...) label the choice of $k_z$ values, with which we calculate the in-plane Wilson loop spectra in Fig.~\ref{CN}.
    }
    \label{Pfa}
\end{figure}

\begin{figure}[t]
    \includegraphics[width=0.24\columnwidth]{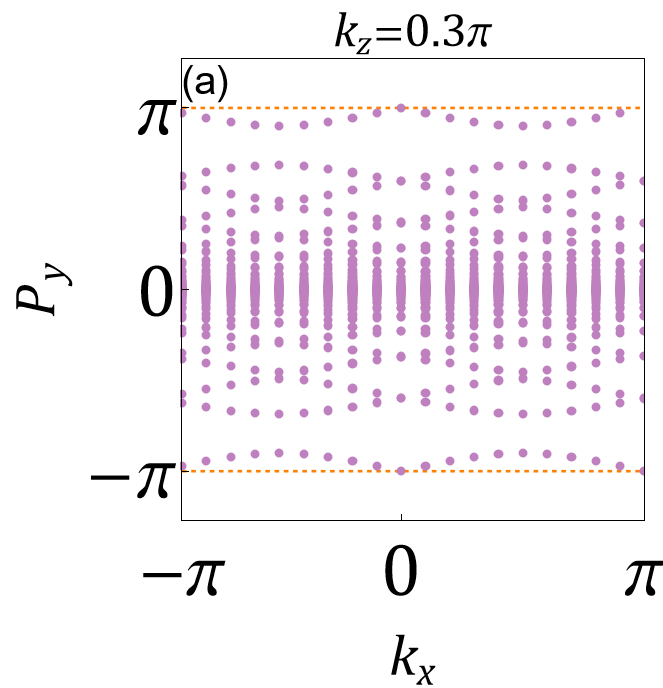}
    \includegraphics[width=0.24\columnwidth]{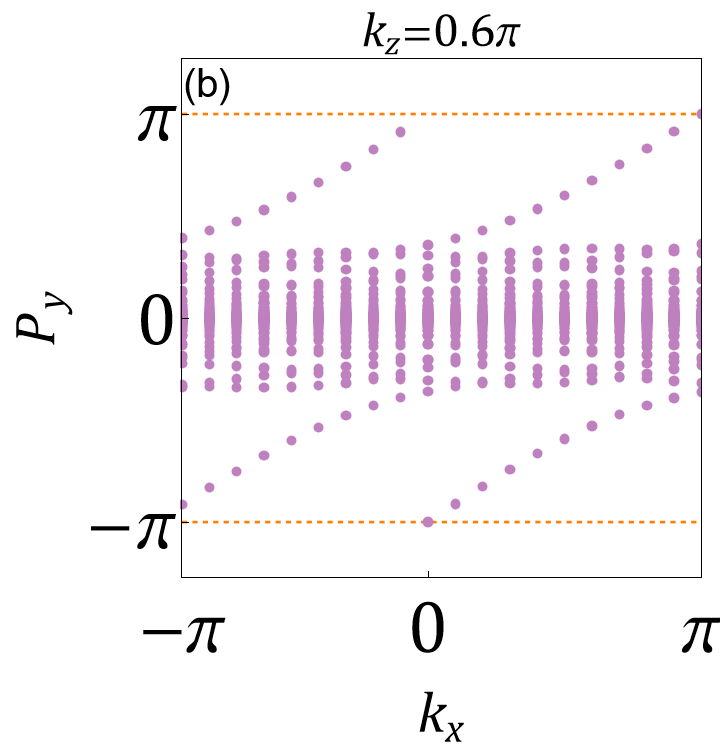}
    \includegraphics[width=0.24\columnwidth]{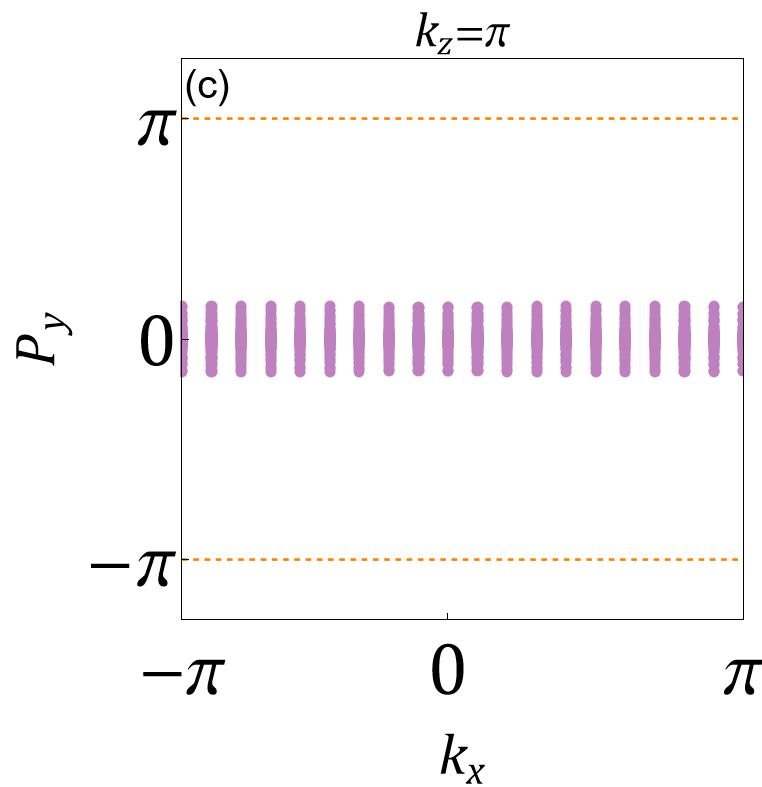}
    \includegraphics[width=0.24\columnwidth]{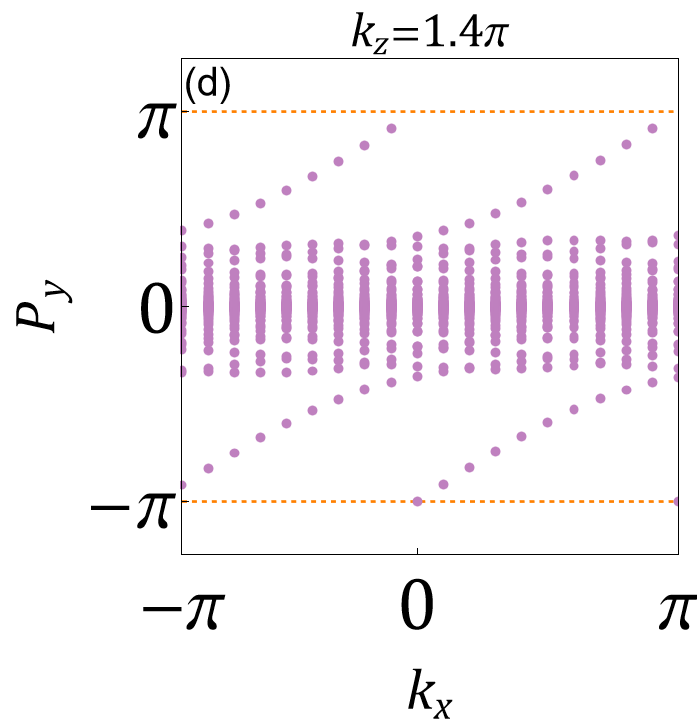}
    \includegraphics[width=0.24\columnwidth]{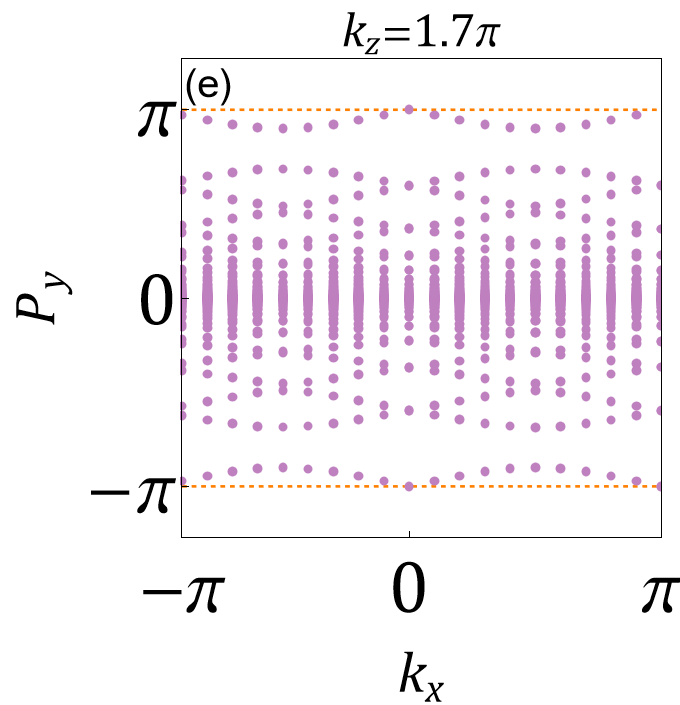}
    \includegraphics[width=0.24\columnwidth]{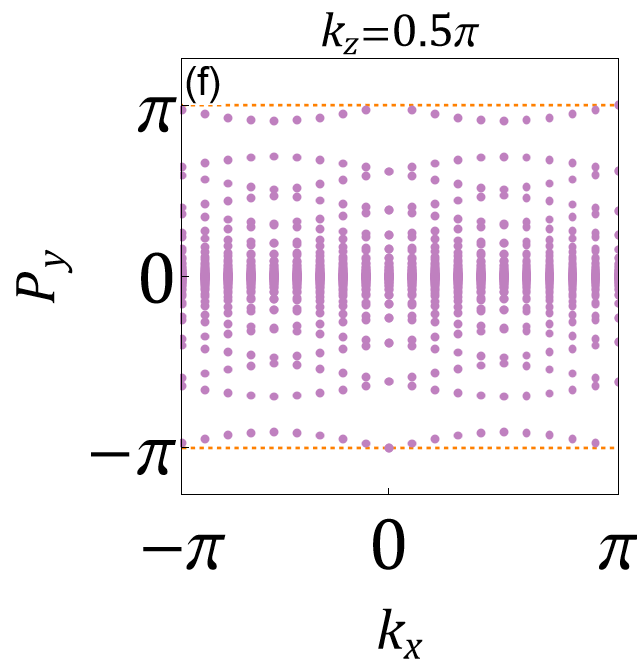}
    \includegraphics[width=0.24\columnwidth]{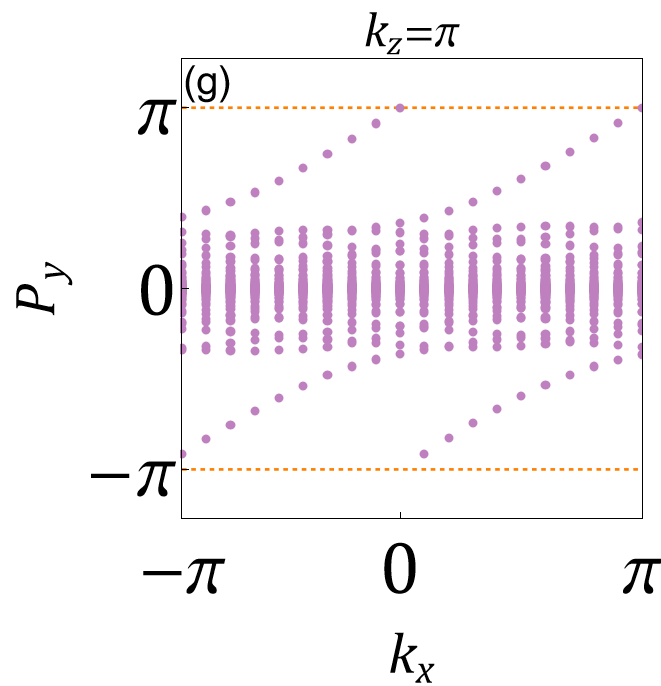}
    \includegraphics[width=0.24\columnwidth]{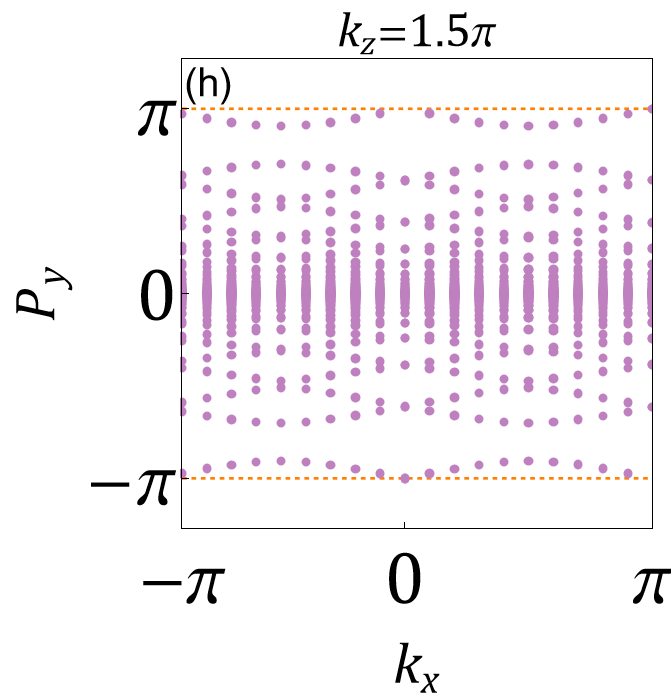}
    \caption{Wilson loops along $k_y$ axis for $M_0=1.8$ in (a-e) with $\nu=2$ and $M_0=1.45$ in (g-h) with $\nu=1$. The value of $k_z$ is fixed for each calculation and is further labeled on top of each figure. The corresponding energy dispersions are shown in Fig.~\ref{Pfa}. We have chosen $L=10$, $M_1=v=2M_2=1$, $\mu=0.1$ and $\Delta_0=0.15$ for all of the calculations.
    }
    \label{CN}
\end{figure}

\subsection{D3. Topological Invariants of CdGM Fermi Surfaces}

Based on the effective theory, we conclude that both the $\mathbb{Z}_2$ topological charge and the Chern number must change across a CdGM Fermi surface. In this part, we explicitly confirm this expectation by numerically calculating the above topological numbers in our microscopic simulations. 

We focus on both $\nu=2$ and $\nu=1$ phases whose energy dispersions are shown in Fig.~\ref{Pfa}, respectively. In addition, the red dashed lines in Fig.~\ref{Pfa} inform the numerical value of ${\rm sgn}({\rm Pf}[\tilde{\cal H}_m (\vec{k})])$, i.e., the sign of the Pfaffian along the $k_z$ axis with $k_x=k_y=0$. Please note that the Pfaffian is defined for the antisymmetrized ``pseudo-Hamiltonian" $\tilde{\cal H}_m$, but not the original Hamiltonian ${\cal H}_m$. Clearly, ${\rm sgn}({\rm Pf}[\tilde{\cal H}_m (\vec{k})])$ flips its sign every time it crosses the Fermi surface. This directly confirms that all of our CdGM Fermi surfaces feature the $\mathbb{Z}_2$ topological charge $\eta=1$.   

We further calculate the value of Chern number ${\cal C}$ for a fixed $k_z$ by evaluating the $k_x$-dependent Wilson loops along the $k_y$ axis. In particular, Figs.~\ref{CN} (a) - (e) show the Wilson loop spectra for the $\nu=2$ phase in Fig.~\ref{Pfa} (a). Similar results for the $\nu=1$ phase in Fig.~\ref{Pfa} (b) are shown in Figs.~\ref{CN} (f) - (h). It is straightforward to see that 
\begin{itemize}
    \item ${\cal C}$ always changes by $2$ when passing through a CdGM Fermi surface.
\end{itemize}
This explicitly justifies the validity of our effective model ${\cal H}$, especially about the in-plane part of ${\cal H}$ resembling two copies of the square-lattice Haldane model. Furthermore, a change of ${\cal C}$ necessarily implies the existence of 3D Weyl nodes that are hidden in the Fermi surface. This agrees with the picture discussed in Ref.~\cite{Agterberg17} that a topologically protected Fermi surface can be viewed as an ``inflation" of point or line nodes.

\section{Appendix E: Size Effect of a Vortex Lattice}

\begin{figure}
    \includegraphics[width=0.24\columnwidth]{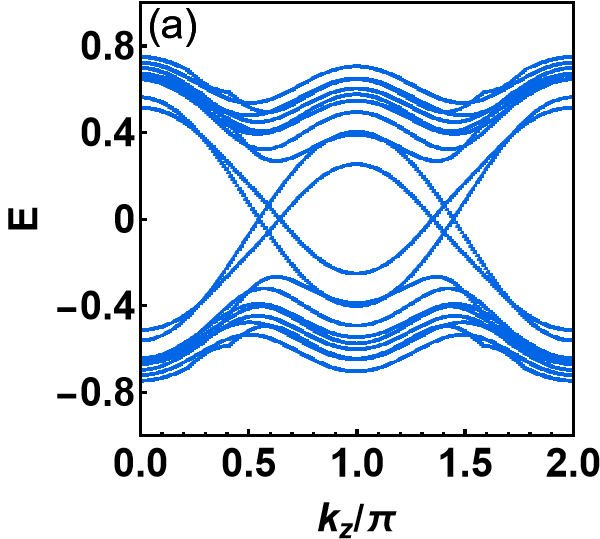}
    \includegraphics[width=0.24\columnwidth]{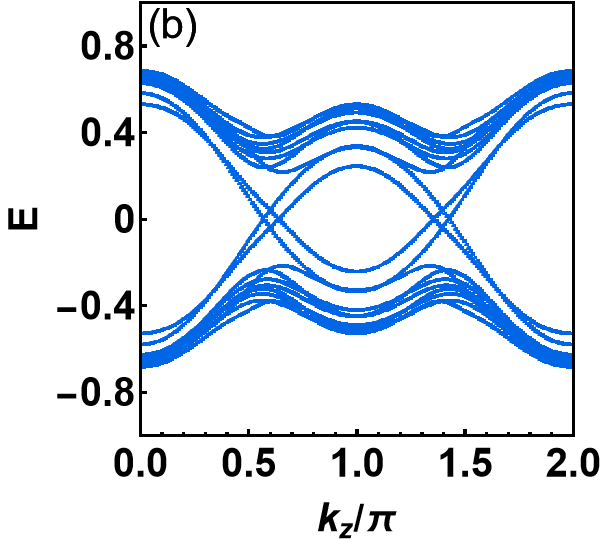}
    \includegraphics[width=0.24\columnwidth]{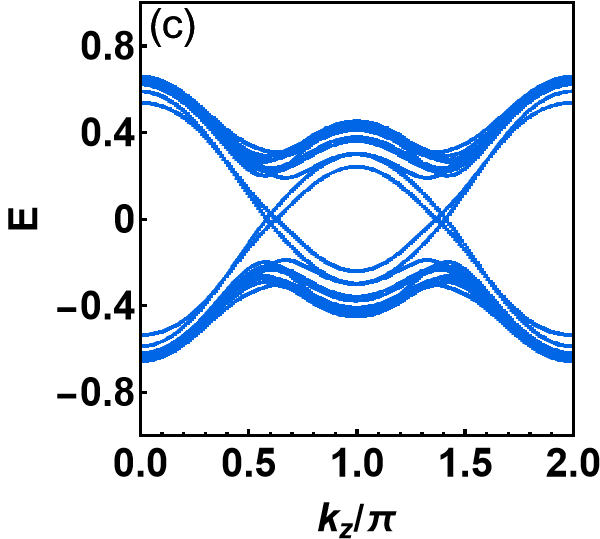}
    \includegraphics[width=0.24\columnwidth]{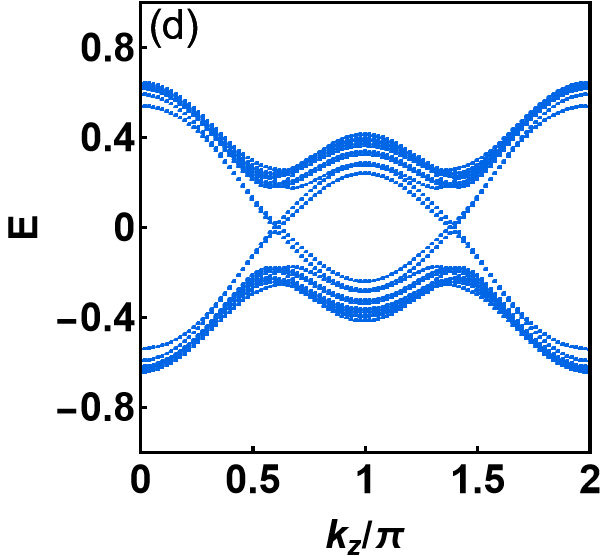}
    \includegraphics[width=0.24\columnwidth]{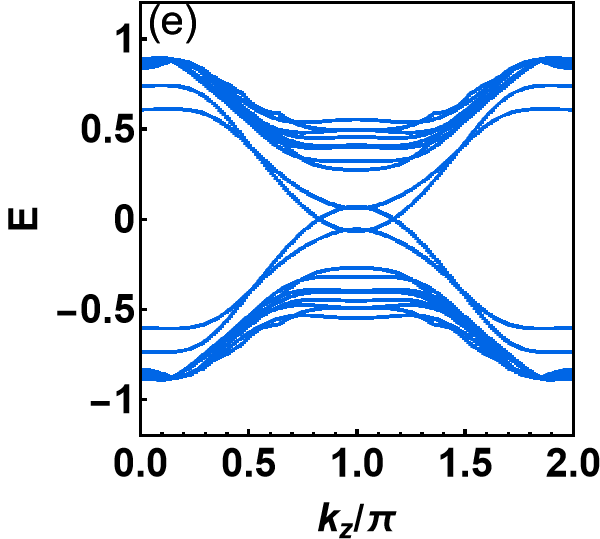}
    \includegraphics[width=0.24\columnwidth]{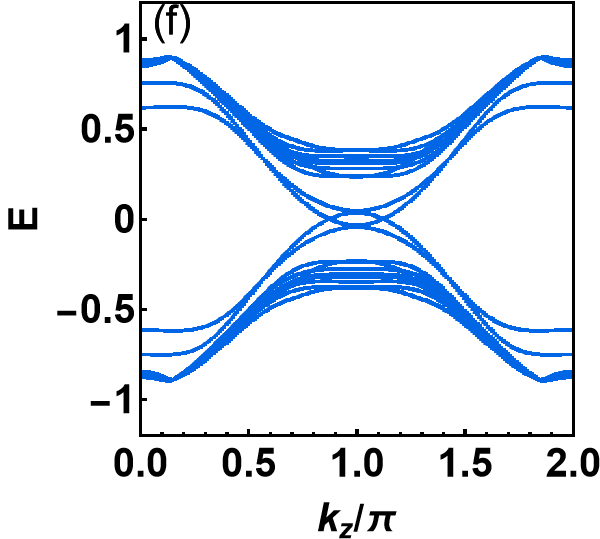}
    \includegraphics[width=0.24\columnwidth]{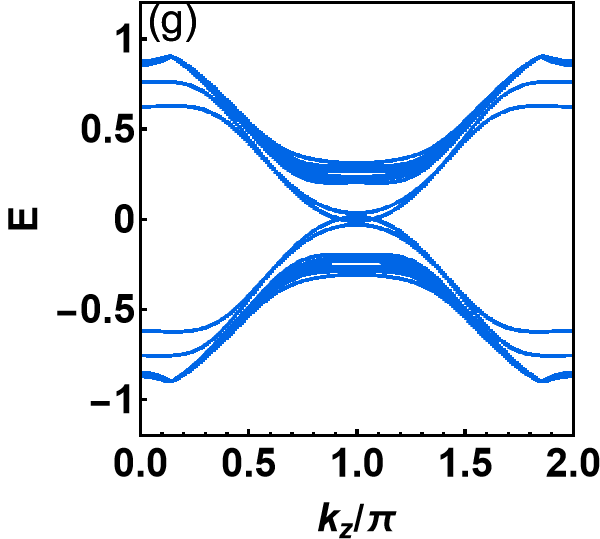}
    \includegraphics[width=0.24\columnwidth]{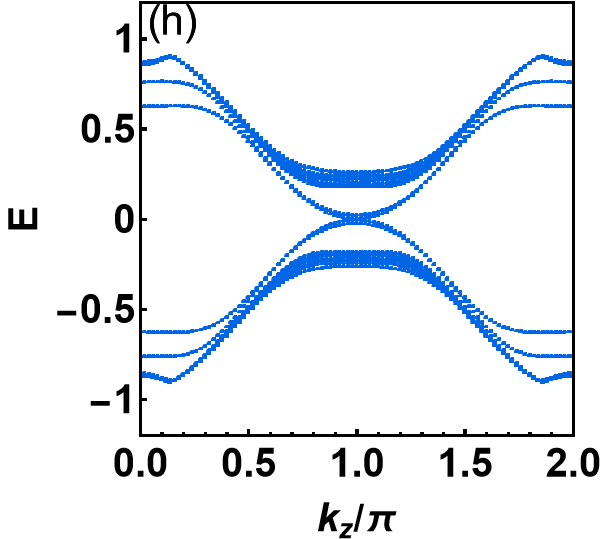}
    \includegraphics[width=0.24\columnwidth]{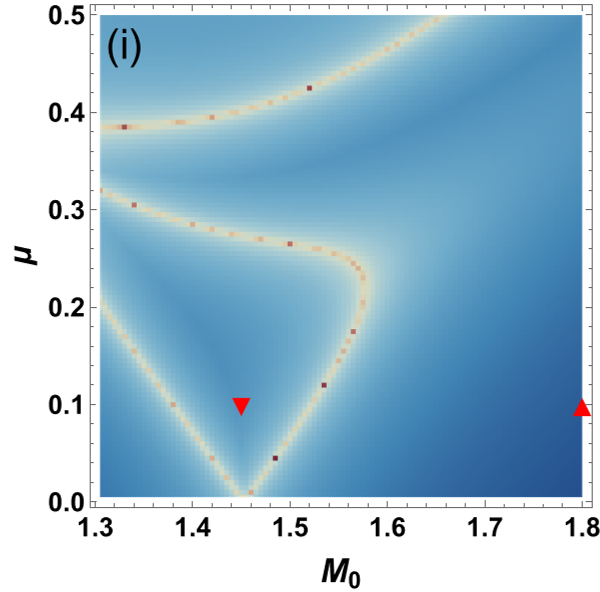}
    \includegraphics[width=0.24\columnwidth]{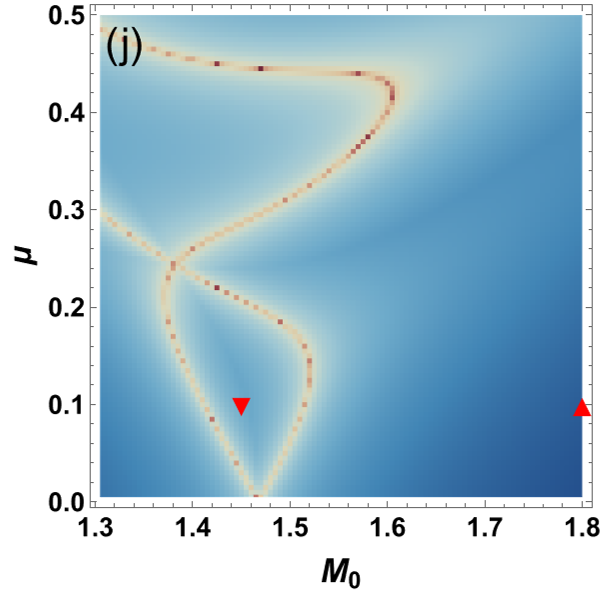}
    \includegraphics[width=0.24\columnwidth]{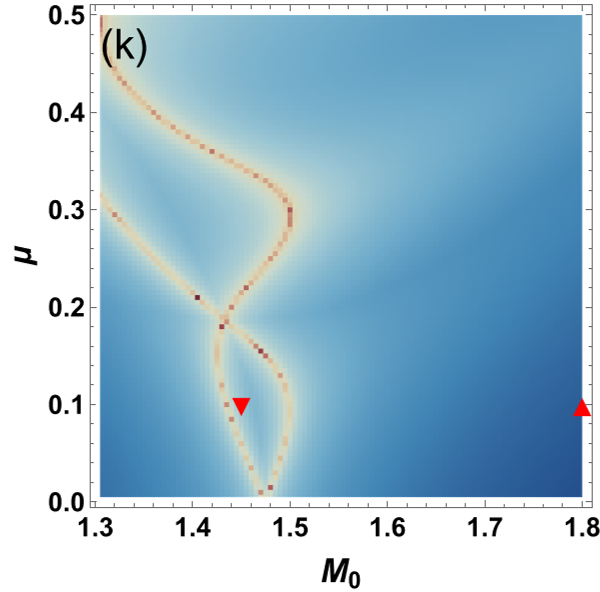}
    \includegraphics[width=0.24\columnwidth]{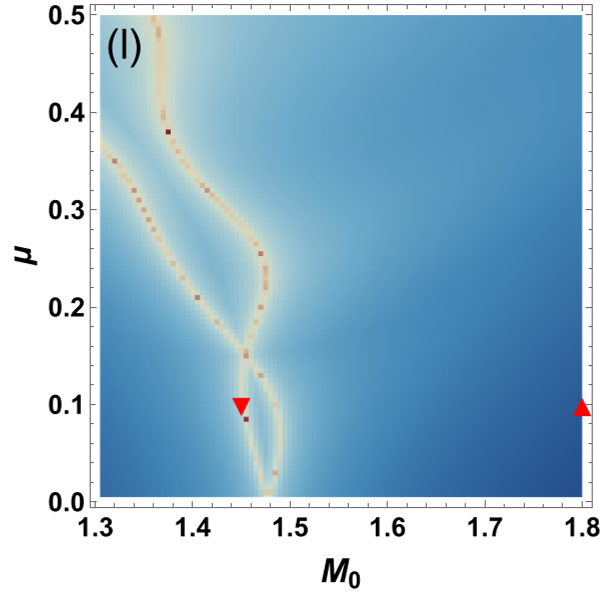}
    \caption{Energy dispersion and phase diagrams as a function of the vortex lattice constant $L$. Each column corresponds to a specific choice of lattice constant $L$. Specifically, we have (a), (e), (i) for $L=10$, (b), (f), (j) for $L=14$, (c), (g), (k) for $L=18$, and (d), (h), (l) for $L=22$. The first row shows energy dispersion along $k_z$ axis for $M_0=1.8,\mu=0.1$, which is also marked by an upward triangle in the corresponding phase diagram in the third row. The second row shows energy dispersion along $k_z$ axis for $M_0=1.45,\mu=0.1$, which is marked by a downward triangle in the corresponding phase diagram in the third row. Other parameters are $M_1=v=2M_2=1$ and $\Delta_0=0.15$.
    }
    \label{KZ_L}
\end{figure}

\begin{figure}
    \includegraphics[width=0.6\columnwidth]{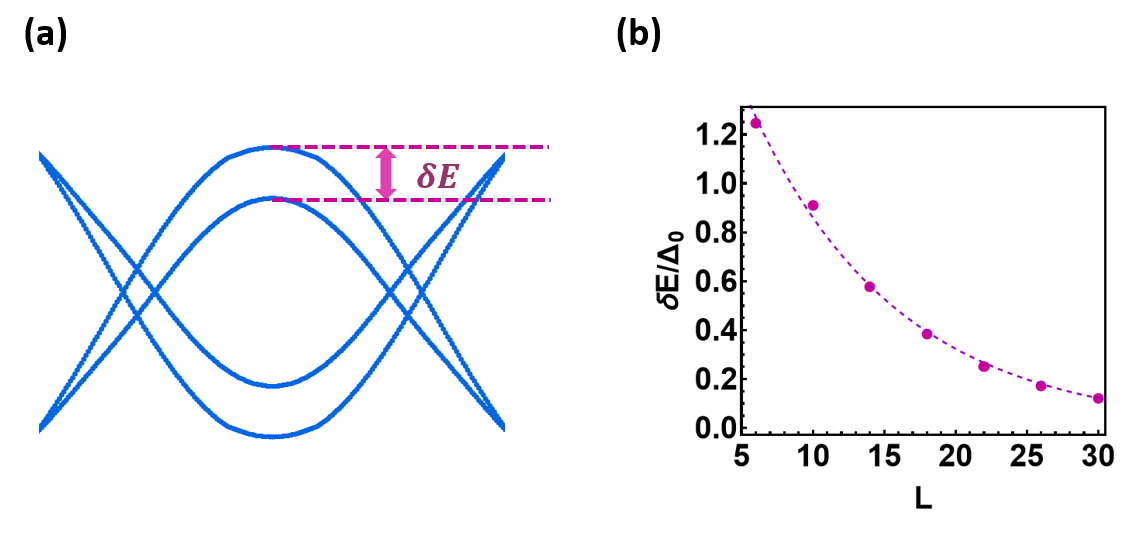}
    \caption{Energy splitting $\delta E$ at $\vec{k}=(0,0,\pi)$ as a function of the vortex lattice constant $L$. We scale the energy by SC pairing gap $\Delta_0=0.15$. The model parameters are $M_1=v=2M_2=1$, $\mu=0.1$, $M_0=1.8$.
    }
    \label{fig4}
\end{figure}

In this appendix, we study how the vortex lattice band structure changes with the vortex lattice constant $L$. All results are obtained from the microscopic lattice model simulation. We first look at the energy dispersion along the $k_z$ axis for a fixed $\mu=0.1$ and a fixed $M_0=1.8$ or $1.45$. The results are shown in Fig.~\ref{KZ_L}. One can see that the level repulsion (between the two lowest bands with $E>0$) becomes smaller when $L$ increases, which is expected as the level repulsion of the lattice hopping should decay exponentially as a function of the lattice constant in general. To confirm this, we plot the level repulsion at $\vec{k}=(0,0,\pi)$ as a function of $L$ in Fig.~\ref{fig4} and find they are well-fitted by 
\begin{eqnarray}
    \frac{\delta E}{\Delta_0} = C_0e^{-\frac{L}{\xi_l}},\ \ \text{with } \xi_l = 10.20,\ C_0 = 2.28. 
\end{eqnarray}

We further check how the topological phase diagrams change as a function of $L$, as shown in Fig.~\ref{KZ_L}. While the quantitative boundaries of the phases vary with $L$, we see the qualitative features of the phase diagram stay the same. 

\section{Appendix F: CdGM Fermi Surface for $\nu=1$ Phase}
\begin{figure}[t]
    \includegraphics[width=0.6\columnwidth]{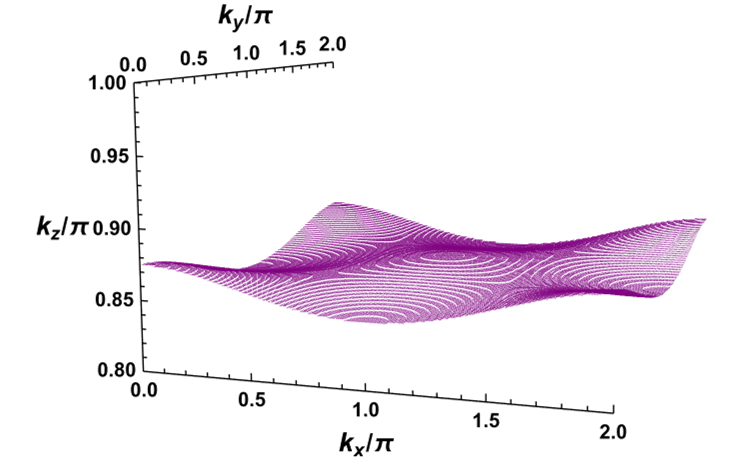}
    \caption{The Fermi surface in the $k_z>0$ half BZ for the $\nu=1$ phase. We use $L=14$, $M_0=1.45$, $\mu=0.1$, $M_1=v=2M_2=1$ and $\Delta_0=0.15$, which are the same as Fig. 3 (e-f) in the main text. 
    }
    \label{FS}
\end{figure}

In Fig. 1 of the main text, we presented the Fermi surface for the $\nu=2$ case. For completeness, we also present the Fermi surface for the $\nu=1$ case, as plotted in Fig.~\ref{FS}. We focus on the $k_z>0$ BZ as there is a $z$-directional mirror symmetry. The parameters we use are $L=14$, $M_0=1.45$, $\mu=0.1$, $M_1=v=2M_2=1$ and $\Delta_0=0.15$, which are the same as Fig. 3 (e-f) in the main text. 

\end{document}